\documentclass[preprint]{revtex4}
\usepackage{amsmath,mathtools,physics,xcolor}
\usepackage{graphicx,multirow}  
\usepackage{float,ragged2e}
\usepackage{booktabs}
\usepackage[verbose,hypertexnames=false]{hyperref}

\hypersetup{colorlinks=false,allbordercolors=blue,pdfborderstyle={/S/U/W 1}}

\begin{document}

\title{Hardy’s nonlocality for entangled pairs in a four-particle system}

\author{Duc Manh Doan}
\affiliation{Nano and Energy Center, Faculty of Physics, VNU University of Science, Vietnam National University, 120401 Hanoi, Vietnam}

\author{Hung Q. Nguyen\footnote{Corresponding author: hungngq@hus.edu.vn.}}

\affiliation{Institute for Quantum Technologies, Technology and Innovation Park, Vietnam National University, Hanoi, Vietnam}
\affiliation{Nano and Energy Center, Faculty of Physics, VNU University of Science, Vietnam National University, 120401 Hanoi, Vietnam}


\begin{abstract}
Nonlocality can be studied through different approaches, such as Bell's inequalities, and it can be found in numerous quantum states, including GHZ states or graph states. Hardy's paradox, or Hardy-type nonlocality, provides a way to investigate nonlocality for entangled states of particles without using inequalities. Previous studies of Hardy's nonlocality have mostly focused on the fully entangled systems, while other entanglement configurations remain less explored. In this work, the system under investigation consists of four particles arranged in a cyclic entanglement configuration, where each particle forms entangled pairs with two neighbors, while non-neighboring particles remain unentangled. We found that this entanglement structure offers a larger set of conditions that lead to the contradiction with the LHV model, compared to the fully entangled systems. This enhancement can be attributed to the presence of multiple excluded states and correlations, in which the measurement result of a particle only influences the result of its paired partners. We implement quantum circuits compatible with the cyclic entanglement structure, and through simulation, the correlation patterns and the states of interest are identified. We further execute the proposed circuits on IBM Brisbane, a practical backend; however, the results show considerable deviations from the simulation counterparts.
\end{abstract}


\maketitle

\section{Introduction}

Nonlocality is a phenomenon in which the measurement statistics of a multipartite quantum system cannot be explained by local realism. This concept was first discussed through the EPR paradox in the article Ref. \cite{einstein1935can}, in which the completeness of quantum mechanics was questioned under the assumption of local realism. Later, in the work Ref. \cite{bell1964einstein}, Bell derived an inequality that must be satisfied by all the LHV (Local Hidden Variables) models. The violation of this inequality by quantum mechanics demonstrated that the LHV model or locality can not reproduce all the quantum predictions. A more widely used version of Bell's inequality was then formulated in the work Ref. \cite{clauser1969proposed} by Clauser, Horne, Shimony, and Holt.

\subsection{Hardy's nonlocality}

Hardy then came up with another approach to demonstrate nonlocality without relying on inequalities. This approach relies solely on the logical contradiction, in which we will be given a set of probability conditions. The contradiction occurs when the LHV model tries to satisfy some of the conditions, but if it does, it will be forced to violate the other. Initially, Hardy's nonlocality is studied through a Gedanken experiment involving electrons and positrons in the work Ref. \cite{PhysRevLett.68.2981}. Later, this proof is extended to a pair of entangled spin-1/2 particles, excluding maximally entangled states, in Ref. \cite{PhysRevLett.71.1665}. These two works are discussed briefly in the following parts.

\subsubsection{The Mach-Zehnder interferometers experiment}

The Mach-Zehnder experiment is a Gedanken experiment, and is the first time Hardy proposed the approach demonstrating nonlocality without inequality. The way this experiment works can be described as follows. Two types of particles in this experiment are positrons and electrons. The experiment consists of two Mach-Zehnder interferometers, $MZ^+$ for positrons and $MZ^-$ for electrons. Initially, those particles are sent through two fixed beam splitters $BS1^{\pm}$. After going through these two beam splitters, the particles' paths are divided into two possible paths, which are $u^\pm$ and $v^\pm$. This experiment is designed so that if the particles follow the path $u^\pm$, they will meet at the point P and annihilate each other. If the particles do not meet, they will approach the removable beam splitters $BS2^\pm$, after which their paths are transformed into $d^\pm$ and $c^\pm$. Finally, they'll reach the corresponding detectors $D^\pm$ and $C^\pm$.

This experiment has several important features. The first thing to note is that the beam splitters $BS1^\pm$ and $BS2^\pm$ can be seen as tools that transform the bases of particles. The $BS1^\pm$ change particles' basis to \{$\ket{u^\pm},\ket{v^\pm}$\}, meanwhile, the $BS2^\pm$ transform them to \{$\ket{c^\pm},\ket{d^\pm}$\}. Following this, in the \{$\ket{u^\pm},\ket{v^\pm}$\} basis, one of the possible outcomes is the annihilation even at P, which corresponds to the state $\ket{u^+}\ket{u^-}$ being excluded. This event directly leads to the first quantum prediction, or the first condition

\begin{equation}
    C_Q^+(\infty)C_Q^-(\infty)=0.
    \label{e1}
\end{equation}
where the $\infty$ denotes that $BS2^\pm$ are removed from particles' paths, meanwhile, the 0 will denote that $BS2^\pm$ are in place.

Subsequently, the purpose of the next two predictions is to observe the influence on a particle when changing the choice of measurement on the other particle. To do this, $BS2^+$ is removed from the positrons' path while $BS2^-$ remains on the electrons' path and vice versa. This leads to the next two quantum predictions

\begin{align}
    &\text{If $D_Q^+(0)=1$ then $C_Q^-(\infty)=1$},\label{e2}\\
    &\text{If $D_Q^-(0)=1$ then $C_Q^+(\infty)=1$}.\label{e3}
\end{align}

The final quantum prediction is obtained when two $BS2^\pm$ are in place, and the state of interest is $\ket{d^+}\ket{d^-}$

\begin{equation}
    D_Q^+(0)D_Q^-(0)=1.
    \label{e4}
\end{equation}

The predictions \eqref{e1}$\rightarrow$\eqref{e4} are seen as the four conditions in Hardy's nonlocality. The question is whether the local realism or the LHV model can reproduce these predictions. The LHV model assumes that the choices of measurement on a particle no longer influence other particles' results. In other words, the particle's results are now solely dependent on a local hidden variable $\lambda$. Based on this model, assume that it can reproduce the last condition \eqref{e4}

\begin{equation}
    D^+(\lambda,0)D^-(\lambda,0)=1.
    \label{e5}
\end{equation}

Following this, with the same variable $\lambda$, this model attempts to reproduce the predictions \eqref{e2}-\eqref{e3} as well

\begin{align}
    &D^+(\lambda,0)C^-(\lambda,\infty)=1,\label{e6}\\
    &D^-(\lambda,0)C^+(\lambda,\infty)=1.\label{e7}
\end{align}

This is the point where the paradox emerges. To satisfy \eqref{e4}, the LHV model already produced \eqref{e5} and matched the corresponding quantum prediction. However, to also satisfy \eqref{e2}-\eqref{e3}, it needs to produce $C^+(\lambda,\infty)=1$ and $C^-(\lambda,\infty)=1$ simultaneously, and this contradicts with the first prediction \eqref{e1}. Therefore, as we can see, the LHV could not reproduce all quantum predictions. It can reproduce some, but eventually, it will be forced to violate the other.

\subsubsection{Hardy's paradox in any two-entangled particle system}

In the case of any two entangled particles, the system is described in two orthonormal bases \{$\ket{u_i},\ket{v_i}$\} and \{$\ket{c_i},\ket{d_i}$\}, where i=1,2. Besides, two non-commuting observables are used, which are $U_i=\ket{u_i}\bra{u_i}$ and $D_i=\ket{d_i}\bra{d_i}$.
Initially, the system is designed to exclude the state $\ket{u_1u_2}$, and eventually, the state equation is described as 
\begin{equation}
    \ket{\psi}=N(\ket{c_1}\ket{c_2}-A^2\ket{u_1}\ket{u_2}).
    \label{e8}
\end{equation}
with $N$ is the normalization constant.

By changing either one or both particles' basis between these two orthonormal bases, four other state equations that are equivalent to Eq.\eqref{e8} are obtained. From those five state equations, there are states of interest that indicate the conditions in Hardy's paradox. Before going to the specific conditions, there is a convention that needs to be clarified. From now on, probabilities having the form $P(A_iB_j=1)$, such as $P(U_1U_2=1)$, will be presented shortly as $P(A_iB_j)$. Back to the main point, in Hardy's original work for two entangled particles \cite{PhysRevLett.71.1665}, the conditions are
\begin{align}
    &P(U_1U_2)=0, \label{e9}\\
    &P(D_1U_2)>0:\text{if $D_1=1$ then $U_2=1$},\label{e10}\\
    &P(U_1D_2)>0:\text{if $D_2=1$ then $U_1=1$},\label{e11}\\
    &P(D_1D_2)>0.\label{e12}
\end{align}

Probability \eqref{e9} can be seen as the first setup for the contradiction with the LHV model. Specifically, the state $\ket{u_1u_2}$ is the quantum prediction that the joint outcome never occurs, which is implied in  Eq.\eqref{e8}. The coefficient of the term $\ket{u_1u_2}$ in Eq.\eqref{e8} is chosen on purpose to exclude this term, just like the annihilation event of the Mach-Zehnder experiment. Meanwhile, two conditions \eqref{e10} and \eqref{e11} show the correlation between two particles. The pattern is that if D = 1 is measured on a particle, then U = 1 has to be obtained on the other. These two conditions are obtained by changing one particle's basis to \{$\ket{c_i},\ket{d_i}$\}, while the other is still in \{$\ket{u_i},\ket{v_i}$\}. The last condition \eqref{e12} is obtained when both particles are in \{$\ket{c_i},\ket{d_i}$\}. In addition, in this work, the last probability condition is referred to as $P_{\rm{success}}$.

\textbf{\textit{The contradiction with the LHV model}} The conditions \eqref{e9}-\eqref{e12} are the quantum predictions, so once again, the question is whether the LHV model can reproduce these quantum predictions. At first, assume that the LHV can satisfy the last condition \eqref{e12}, which means
\begin{equation}
    D_1(\lambda)D_2(\lambda)=1.
    \label{e13}
\end{equation}

After that, with the same local hidden variable $\lambda$, it also attempts to reproduce the predictions \eqref{e10}-\eqref{e11}
\begin{align}
    &D_1(\lambda)U_2(\lambda)=1,\label{e14}\\
    &U_1(\lambda)D_2(\lambda)=1.\label{e15}
\end{align}

As can be seen from \eqref{e14} - \eqref{e15}, in an attempt to satisfy the two correlation conditions \eqref{e10}-\eqref{e11}, the LHV model inevitably causes $U_1(\lambda)U_2(\lambda)=1$, which directly contradicts the first condition \eqref{e9}. The paradox still emerges just like in the Mach-Zehnder experiment, where the LHV model couldn't reproduce all of the quantum predictions. The LHV model can satisfy some of the conditions, but if it does, it will be forced to violate the other conditions. The contradiction between the LHV model and quantum predictions above is the heart of Hardy's nonlocality. Furthermore, it is pointed out in Ref. \cite{PhysRevLett.71.1665} that the nonlocal effect is maximum when $P(D_1D_2)$ or $P_{\rm{success}}$ is maximum. In other words, the nonzero probability of $\ket{d_1d_2}$ is vital in every demonstration of Hardy's nonlocality, and the higher this value is, the more visibly the nonlocal effect can be observed.

\subsection{A stronger Hardy's nonlocality}

In the two previous parts, the original Hardy's nonlocality was discussed. As we can see, the state $\ket{d_1d_2}$ is necessary to cause the paradox, because its appearance, together with the two correlation conditions, form a trap forcing the LHV model to violate the first condition. This raises a question whether other states in the \{$\ket{c_i},\ket{d_i}$\} basis can cause the paradox as well. In the work Ref. \cite{PhysRevA.107.042210}, it was found that there are other states besides the standard state $\ket{d_1d_2...d_n}$ that can also make the paradox emerge. Accordingly, with their finding, the $P_{\rm{success}}$ is now larger than that in the standard cases, which means that the nonlocality effect is stronger as well. The probability $P_{\rm{success}}$ is now the sum of the probability of all states that can cause the contradictions, not just $P(D_1D_2...D_n)$.

Specifically, in Ref. \cite{PhysRevA.107.042210}, they pointed out that in an $n>2$ fully entangled system, states that have at least two particles simultaneously yielding D=1 are enough to cause the Hardy paradox. For instance, with n=3, based on this approach, there will be in total four states that can create the contradiction, which are $\ket{d_1d_2c_3}, \ket{d_1c_2d_3}, \ket{c_1d_2d_3},\ket{d_1d_2d_3}$. Let's have a test with $\ket{d_1d_2c_3}$. Similar to Hardy's original work, the set of  conditions in this case is

\begin{align}
    &P(U_1U_2U_3)=0, \label{e16}\\
    &P(D_1U_2U_3)>0:\text{if $D_1=1$ then $U_2=U_3=1$},\label{e17}\\
    &P(U_1D_2U_3)>0:\text{if $D_2=1$ then $U_1=U_3=1$},\label{e18}\\
    &P(U_1U_2D_3)>0:\text{if $D_3=1$ then $U_1=U_2=1$},\label{e19}\\
    &P(D_1D_2C_3)>0.\label{e20}
\end{align}

If the LHV model can satisfy condition \eqref{e20} with the local variable $\lambda$, then
\begin{equation}
    D_1(\lambda)D_2(\lambda)C_3(\lambda)=1.
    \label{e21}
\end{equation}

To also reproduce the correlation conditions \eqref{e17} - \eqref{e18}, the LHV model needs to cause $U_2=U_3=1$ and $U_1=U_3=1$. This still inevitably leads to $U_1(\lambda)U_2(\lambda)U_3(\lambda)=1$, which violated the first condition \eqref{e16}. Accordingly, instead of having just one state $\ket{d_1d_2d_3}$ to cause the contradiction, we now have three more states that can do the same thing. Additionally, instead of $P_{\rm{success}}=P(\ket{d_1d_2d_3})$ like in the original approach, $P_{\rm{success}}$ is now determined as the sum of probabilities of states that can cause the paradox. In the 3-fully entangled particle system, this value is determined as

\begin{equation}
        P_{success}=P(D_1D_2C_3)+P(D_1C_2D_3)+P(C_1D_2D_3)+P(D_1D_2D_3).
        \label{e22}
\end{equation}

The selection rule for these states depends on the correlation among particles. In other words, in different entanglement configurations, a different chosen rule is used. The chosen rule in  Ref. \cite{PhysRevA.107.042210} is for a fully entangled system; applying it to other entanglement configurations might not form the paradox as desired. Additionally, with this approach, $P_{\rm{success}}$ also implies that there is a state, or more than one state, that can cause the contradiction between quantum predictions and locality.

Another notable point of Ref. \cite{PhysRevA.107.042210} is that they were able to generalize Hardy-type nonlocality for n fully entangled particles, and realize Hardy's experiment on quantum computers as well. In accordance with Hardy's work \cite{PhysRevLett.71.1665}, to obtain the first condition, the coefficient of the state $\ket{u_1u_2...u_n}$ is chosen on purpose so that it'll be excluded when represented in the \{$\ket{u_i},\ket{v_i}$\} basis. However, this method becomes quite challenging when increasing the system's size. In \cite{PhysRevA.107.042210}, The state equation for the first condition relates to its quantum circuit. In particular, the n qubits are initially unentangled, and after they undergo a Toffoli gate, where all of them are control qubits, they become entangled, and the state $\ket{u_1u_2...u_n}$ is excluded as well. This process is referred to as the post-selection process. The state equation is constructed based on this post-selection process, and it's shown as follows 

\begin{equation}
    \ket{\psi}=N[\ket{c_1c_2...c_n}-\prod_{k=1}^nA_k\ket{u_1u_2...u_n}].
    \label{e23}
\end{equation}

The first term of this equation comprises all possible states of the system when described in the \{$\ket{u_i},\ket{v_i}$\} basis. This term represents the system at the beginning, where the qubits are unentangled. After the Toffoli gate, the state $\ket{u_1u_2...u_n}$ is excluded, and this process is represented simply by subtracting this state from $\ket{c_1c_2...c_n}$. This equation is quite similar to Eq.\eqref{e8}, but it describes the results of the quantum circuits precisely. 

\subsection{Other approach to study Hardy's nonlocality}

Hardy's nonlocality can be proved for any choice of two different measurement possibilities \cite{PhysRevA.50.62} and can be extended to a system of n-particles \cite{PhysRevA.56.1023}, where it was demonstrated that all non-product states admit Hardy's nonlocality. Further exploration shows that this type of nonlocality could appear for almost all entangled states of three spin-1/2 particles under certain conditions relating to the states' coefficients \cite{WU1996129}. For any entangled states and any maximally entangled states, a system of three spin-1/2 particles exhibits Hardy-type nonlocality with the maximum value of $P_{\rm{success}}\approx12.5\%$ \cite{GHOSH1998249}. 

In addition, the work Ref. \cite{WU2000221} showed that GHZ states can be used to reject the local hidden variables model under ideal conditions. It is then expanded to generalized n-particle GHZ states in Ref. \cite{CERECEDA2004433}. Subsequently, it is extended, and higher values of $P_{\rm{success}}$ for $n\geq3$ are found \cite{PhysRevLett.120.050403, CERECEDA2004433}. Furthermore, the Hardy-type nonlocality is proved for graph states in the works Ref.  \cite{cabello2008nonlocality} and \cite{PhysRevLett.116.070401}. Hardy-type nonlocality has also been demonstrated for other states such as W and Dicke states \cite{PhysRevA.91.032108} or symmetric states \cite{PhysRevLett.108.210407}. 

The comparison of Hardy nonlocality with Bell's inequality or CHSH inequality can be found in Ref. \cite{PhysRevA.52.2535}, \cite{PhysRevA.78.032114}, \cite{GHIRARDI20081982}, \cite{1468301}. In addition, the relationship between nonlocality and entanglement \cite{PhysRevLett.95.210402, Brunner_2005, PhysRevA.83.022108, PhysRevA.97.062313} has also attracted significant interest. Experimental demonstrations for nonlocality have been reported using atomic systems  \cite{PhysRevLett.100.150404,doi:10.1126/science.1221856} and photonic platforms \cite{yokota2009direct, PhysRevLett.102.020404,luo2018experimental}.

Quantum computers have emerged as useful tools for studying quantum phenomena, especially for simulating physical systems \cite{PhysRevD.108.023013, PhysRevX.6.031007, doi:10.1137/18M1231511, PhysRevX.8.031022, kokail2019self, martinez2016real} or machine learning \cite{havlivcek2019supervised, cong2019quantum, farhi2018classification}. Within the study of quantum nonlocality, executing quantum circuits on quantum computers offers an alternative observation of particle correlations and opportunities for examining different entangled states. Specifically, in the work Ref. \cite{PhysRevLett.120.050403}, quantum computers are employed to quantify the threshold visibility, which is the purity required for a quantum state to exhibit nonlocality under realistic noise. 

\subsection{Hardy's nonlocality in the cyclic entanglement}

Based on the quantum circuits construction demonstrated in Ref. \cite{PhysRevA.107.042210}, it is possible to modify the entanglement configuration of the system by changing the setup of the Toffoli gates. This naturally raises a question of whether Hardy's nonlocality would behave the same when the entanglement structure is altered. 

In this work, we modified the Toffoli gates setup in a four-qubit system to produce a different entanglement configuration. Instead of using a single Toffoli gate connecting all qubits, we implement multiple distinct Toffoli gates, each acting on two neighboring qubits as control qubits. These gates are interconnected through shared qubits, forming a closed loop. Therefore, the corresponding entanglement configuration can be seen as the cyclic entanglement structure, also known as the ring cluster. In physical terms, each particle forms entangled pairs with its neighbors; meanwhile, the two non-neighbors remain unentangled.

The four-particle system is chosen for examination because it requires $n\geq4$ for the correlation of cyclic configuration to be clearly distinguished from the fully entangled cases. If $n\leq3$, the particle's correlations can resemble those of fully entangled cases. 

\section{Method}
\subsection{Conceptual}

As discussed in the first section, there are three main goals of the conditions. Based on these goals, the conditions are divided into three sets.

The first set of conditions can be viewed as the first setup for the later contradiction, as the joint probabilities in this set are often those that the LHV would be forced to contradict. To obtain the conditions in the first set, the system is designed to exclude specific states. Those states are often chosen based on the entanglement configuration or the way that the path $u_k (k=1,2,...,n)$ of particles crosses each other. For instance, the state $\ket{u_1u_2}$ in the standard case corresponds to fully entangled systems, and in terms of the optic experiment, it corresponds to the paths $u_1$ and $u_2$ of two particles intersecting.

For the second set of conditions, the main goal is to identify the correlations among the particles. The approach in forming these conditions is the same as Ref. \cite{PhysRevLett.71.1665} and  \cite{PhysRevA.107.042210}. In particular, one particle's basis is changed to another basis while the rest remain the same. This process will be done for each particle one by one. Accordingly, the number of state equations and quantum circuits in this second set is equal to the number of particles.

In terms of the third set of conditions, all particles' basis are transformed to another basis. Based on the correlations among particles, several states of interest are chosen. They are chosen so that their corresponding joint probabilities, together with the second set, form a trap for the LHV model to force it to violate some conditions in the first set. Additionally, the $P_{\rm{success}}$ is determined as the sum of the probability of those states of interest.

\subsection{Analogy between quantum circuit and optical experiment}

As mentioned in the previous part, changing the Toffoli gates setup can change the entanglement configuration as well. Therefore, one of the main analogies between the quantum circuit and the optical setup is the role of the Toffoli gate and the intersection point P, where the path $u_k (k=1,2,...,n)$ crosses. In the fully entangled, according to \cite{PhysRevA.107.042210}, a single Toffoli gate connecting all qubits is similar to all the paths $u_k (k=1,2,..n)$ crossing each other at a point P. The results are all the same, which is the exclusion of the state $\ket{u_1u_2...u_n}$.

\begin{figure}[ht]
    \centering \includegraphics[width=1\textwidth,keepaspectratio]{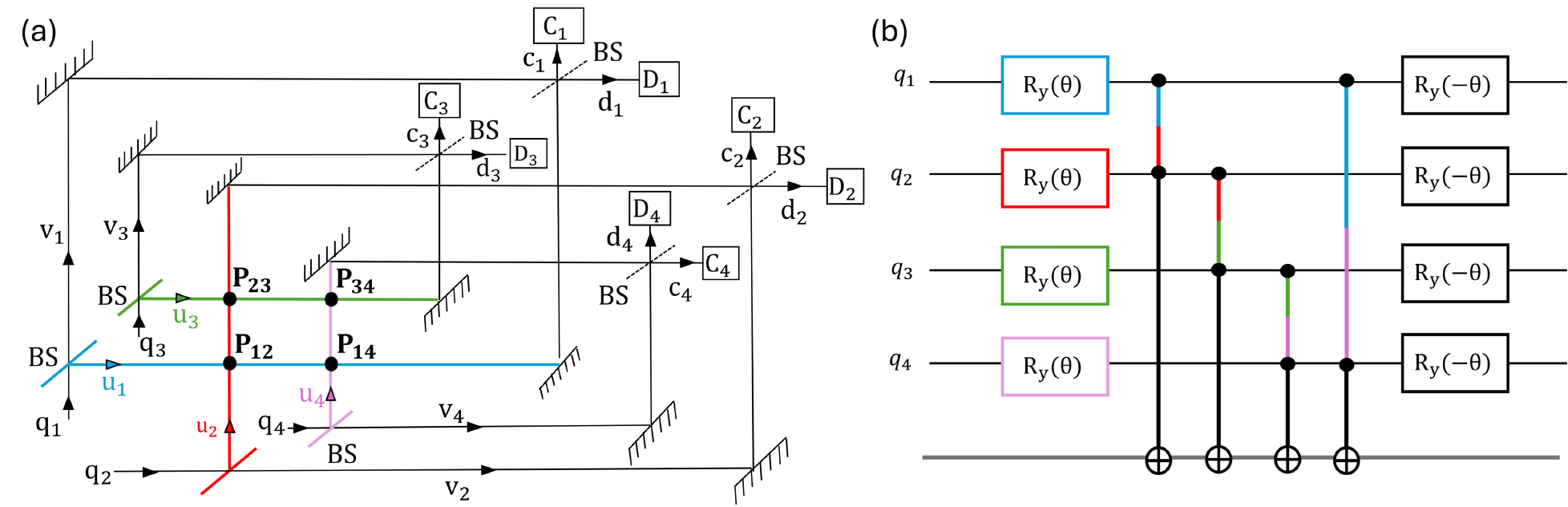}
    \caption{(a) Four Mach-Zehnder interferometers overlapping in pairs. Each particle travels along two possible paths labeled $v_i$ and $u_i$. The $u_i$ paths, highlighted in different colors, are trajectories where particles meet and annihilate each other at points $P_{ij}$. Specifically, $P_{12}$ corresponds to the intersection of $u_1$ and $u_2$; similarly, $P_{23}$ corresponds to $u_2$ and $u_3$, $P_{34}$ corresponds to $u_3$ and $u_4$, and $P_{14}$ corresponds to $u_1$ and $u_4$. Besides, fixed beam splitters are highlighted in different colors. (b) The quantum circuit setup of (a), four rotation gates $R_y(\theta)$ are highlighted to match the fixed beam splitters in the optical setup, which divide the particles' path into $u_i$ and $v_i$. Moreover, the line connecting two control qubits of Toffoli gates indicates that these gates entangle two qubits, analogous to the intersection of two particles' paths.}
    \label{1}
\end{figure}

In terms of the cyclic entanglement configuration, the Toffoli gate setup is that multiple Toffoli gates are used in the circuit, each of which uses two neighboring qubits as control qubits, and these gates are connected by shared qubits. Since in this work, we focus on the four-particle system, there will be four Toffoli gates in the quantum circuits, and the analogy with the optical setup is shown in Fig. \ref{1}.

Fig. \ref{1} is our assumption for the analogy between the optical experiment and the corresponding quantum circuit for the cyclic entanglement. The four Toffoli gates in Fig. \ref{1}(b) are analogous to four intersection points $P_{ij}(i,j=1,2,3,4)$. In terms of the optical setup, after each $P_{ij}$, states having the pattern $\ket{..u_iu_j...}$ are excluded. For instance, after $P_{12}$, states $\ket{u_1u_2x_3y_4}$ are excluded, where x, y can be u or v. Therefore, after going through the four Toffoli gates, the pattern of the excluded states is that at least two particles of the same entangled pair have U=1. It's expected to have nine such states, and from now on, these states are referred to as vanished states

\begin{equation}
    \begin{split}
         \{&\ket{u_1u_2u_3u_4}, \ket{u_1u_2u_3v_4}, \ket{u_1u_2v_3u_4}, \ket{u_1u_2v_3v_4}, \ket{v_1u_2u_3v_4},\\& \ket{v_1u_2u_3u_4}, \ket{v_1v_2u_3u_4}, \ket{u_1v_2u_3u_4}, \ket{u_1v_2v_3u_4}\}. 
         \label{e24}
    \end{split}
\end{equation}

\subsection{Mathematical description}

The state equations of each set of conditions are described in two orthonormal bases \{$\ket{u_k},\ket{v_k}$\} and \{$\ket{c_k},\ket{d_k}$\}. Their relations obey
\begin{align}
    \ket{u_k}=A^*_k\ket{c_k}-B_k\ket{d_k},\label{e25}\\ \ket{v_k}=B^*_k\ket{c_k}+A_k\ket{d_k},\label{e26}
\end{align}
with inverse relations
\begin{align}
    \ket{c_k}&=A_k\ket{u_k}+B_k\ket{v_k},\label{e27}\\
    \ket{d_k}&=-B^*_k\ket{u_k}+A^*_k\ket{v_k},\label{e28}
\end{align}
where k=1,2,3,4. There will be several states of interest in each state equation, and these states imply the corresponding probability conditions in that set.

In addition, two pairs of non-commuting observables are used, which are $U_k=\ket{u_k}\bra{u_k}$, $V_k=\ket{v_k}\bra{v_k}=I - U_k$, and $D_k=\ket{d_k}\bra{d_k}$, $C_k=\ket{c_k}\bra{c_k}=I - D_k$.

\subsubsection{The first set of conditions}

Based on the approach used in Ref. \cite{PhysRevA.107.042210}, the state equation of the first set can be obtained easily by subtracting nine vanished states from the state $\ket{c_1c_2c_3c_4}$. The coefficients of these states are equal to their coefficients in the representation of $\ket{c_1c_2c_3c_4}$ in the \{$\ket{u_k},\ket{v_k}$\} basis

\begin{equation}
    \begin{split}
        \ket{\psi}&=N[\ket{c_1c_2c_3c_4}-A_1A_2A_3A_4\ket{u_1u_2u_3u_4}-A_1A_2A_3B_4\ket{u_1u_2u_3v_4}\\&\quad-A_1A_2B_3A_4\ket{u_1u_2v_3u_4}-A_1A_2B_3B_4\ket{u_1u_2v_3v_4}-B_1A_2A_3B_4\ket{v_1u_2u_3v_4}\\&\quad-B_1A_2A_3A_4\ket{v_1u_2u_3u_4}-B_1B_2A_3A_4\ket{v_1v_2u_3u_4}-A_1B_2A_3A_4\ket{u_1v_2u_3u_4}\\&\quad-A_1B_2B_3A_4\ket{u_1v_2v_3u_4}].
        \label{e29}
    \end{split}
\end{equation}

By substituting Eq.\eqref{e27} and \eqref{e28} into the first term of  Eq.\eqref{e29}, the description of the system in the  \{$\ket{u_k},\ket{v_k}$\} basis will be obtained
\begin{equation}
    \begin{split}
    \ket{\psi}&=N[B_1A_2B_3A_4\ket{v_1u_2v_3u_4}+B_1B_2B_3A_4\ket{v_1v_2v_3u_4}+A_1B_2A_3B_4\ket{u_1v_2u_3v_4}\\&\quad+B_1B_2A_3B_4\ket{v_1v_2u_3v_4}+A_1B_2B_3B_4\ket{u_1v_2v_3v_4}+B_1A_2B_3B_4\ket{v_1u_2v_3v_4}\\&\quad+B_1B_2B_3B_4\ket{v_1v_2v_3v_4}].
    \label{e30}
    \end{split}
\end{equation}

where N is the normalization constant. The value of N is calculated as follows, which is shown in detail in the appendix $N=\frac{1}{\sqrt{1+C}},$
{\small
\begin{equation}
    \begin{split}
        C&=-\prod_{i=1}^{4}\lvert A_i \rvert^2-\prod_{i=1}^{3}\lvert A_i \rvert^2\lvert B_4 \rvert^2-\prod_{\substack{i=1 \\ i \neq 3}}^{4}\lvert A_i \rvert^2\lvert B_3 \rvert^2-\prod_{i=1}^{2}\prod_{j=3}^{4}\lvert A_i \rvert^2\lvert B_j \rvert^2-\prod_{\substack{i=1 \\ i \neq 2,3}}^{4}\prod_{j=2}^{3}\lvert B_i \rvert^2\lvert A_j \rvert^2\\&-\lvert B_1 \rvert^2\prod_{i=2}^{4}\lvert A_i \rvert^2-\prod_{i=1}^{2}\prod_{j=3}^{4}\lvert B_i \rvert^2\lvert A_j \rvert^2-\prod_{\substack{i=1 \\ i \neq 2}}^{4}\lvert A_i \rvert^2\lvert B_2 \rvert^2-\prod_{\substack{i=1 \\ i \neq 2,3}}^{4}\prod_{j=2}^{3}\lvert A_i \rvert^2\lvert B_j \rvert^2.
        \label{e31}
    \end{split}
\end{equation}
}
With the nine vanished states in the state equation, there will be nine corresponding probability conditions in this first set. This number is considerably larger than that in the fully entangled cases.

\subsubsection{The second set of conditions}

Regarding the second set, there are four state equations corresponding to changing four qubits' basis one by one. As discussed in previous parts, one qubit's basis is transformed to \{$\ket{c_k},\ket{d_k}$\}, while the remaining qubits are still measured in \{$\ket{u_k},\ket{v_k}$\}. In particular, consider the qubit $k^{th}$, Eq.\eqref{e27} and \eqref{e28} is substituted into the first term of Eq.\eqref{e29} for all particles except $k^{th}$ and Eq.\eqref{e25}, \eqref{e26} are substituted into the minus terms but only for the $k^{th}$ particle. For instance, changing the first qubit's basis (k=1) to \{$\ket{c_k},\ket{d_k}$\} basis

{\small
\begin{equation}
    \begin{split}
        \ket{\psi}&=N[\ket{c_1}\otimes(A_2\ket{u_2}+B_2\ket{v_2})\otimes(A_3\ket{u_3}+B_3\ket{v_3})\otimes(A_4\ket{u_4}+B_4\ket{v_4})\\&-A_1A_2A_3A_4(A^*_1\ket{c_1}-B_1\ket{d_1})\otimes\ket{u_2u_3u_4}-A_1A_2A_3B_4(A^*_1\ket{c_1}-B_1\ket{d_1})\otimes\ket{u_2u_3v_4}\\&-A_1A_2B_3A_4(A^*_1\ket{c_1}-B_1\ket{d_1})\otimes\ket{u_2v_3u_4}-A_1A_2B_3B_4(A^*_1\ket{c_1}-B_1\ket{d_1})\otimes\ket{u_2v_3v_4}\\&-B_1A_2A_3B_4(B^*_1\ket{c_1}+A_1\ket{d_1})\otimes\ket{u_2u_3v_4}-B_1A_2A_3A_4(B^*_1\ket{c_1}+A_1\ket{d_1})\otimes\ket{u_2u_3u_4}\\&-B_1B_2A_3A_4(B^*_1\ket{c_1}+A_1\ket{d_1})\otimes\ket{v_2u_3u_4}-A_1B_2A_3A_4(A^*_1\ket{c_1}-B_1\ket{d_1})\otimes\ket{v_2u_3u_4}\\&-A_1B_2B_3A_4(A^*_1\ket{c_1}-B_1\ket{d_1})\otimes\ket{v_2v_3u_4}],
        \label{e32}
    \end{split}
\end{equation}
}

{\small
\begin{equation}
    \begin{split}
        \Rightarrow\ket{\psi}&=N[\lvert B_1 \rvert^2A_2B_3A_4\ket{c_1u_2v_3u_4}+\lvert B_1 \rvert^2B_2B_3A_4\ket{c_1v_2v_3u_4}+\lvert B_1 \rvert^2A_2B_3B_4\ket{c_1u_2v_3v_4}\\&\quad+B_2A_3B_4\ket{c_1v_2u_3v_4}+B_2B_3B_4\ket{c_1v_2v_3v_4}+A_1B_1A_2B_3B_4\ket{d_1u_2v_3v_4}\\&\quad+A_1B_1A_2B_3A_4\ket{d_1u_2v_3u_4}+A_1B_1B_2B_3A_4\ket{d_1v_2v_3u_4}].
        \label{e33}
    \end{split}
\end{equation}
}
Among eight possible outcomes, as exhibited in Eq.\eqref{e33}, attention should be paid to three terms, which are $\ket{d_1u_2v_3v_4}$, $\ket{d_1v_2v_3u_4}$, and $\ket{d_1u_2v_3u_4}$. The reason is that these states indicate the correlations among particles' measurement results. Specifically, the measurement outcome for the first particle implies that if $D_1=1$, then it causes $U_2=1$, or $U_4=1$, or $U_2=U_4=1$ simultaneously. From these results, a difference with the standard cases has emerged. In the fully entangled configurations, with n = 4 and k=1 as well, the state of interest is only $\ket{d_1u_2u_3u_4}$. However, in this cyclic entanglement, the number of states of interest is now three, and these states imply that a particle can only influence the results of its paired partners, or, in other words, its entangled partners. Using the same method with other particles, this pattern can also be observed. Consequently, in this second set, after changing the basis of four particles one by one, there can be up to twelve correlation conditions.

\subsubsection{The third set of conditions}

The third set's state equation is observed in the \{$\ket{c_k},\ket{d_k}$\}, to do this, all particles are transformed to this basis. Specifically, Eq.\eqref{e25} - \eqref{e26} are substituted into all minus terms of Eq.\eqref{e29}

\begin{equation}
    \begin{split}
        \ket{\psi}&=N[...-\lvert B_1 \rvert^2\lvert B_2\rvert^2A_3B_3A_4B_4\ket{c_1c_2d_3d_4}+...-\lvert B_1 \rvert^2A_2B_2A_3B_3\lvert B_4\rvert^2\ket{c_1d_2d_3c_4}\\&-\lvert B_1 \rvert^2A_2B_2A_3B_3A_4B_4\ket{c_1d_2d_3d_4}-...-A_1B_1\lvert B_2 \rvert^2\lvert B_3 \rvert^2A_4B_4\ket{d_1c_2c_3d_4}\\&-...-A_1B_1\lvert B_2 \rvert^2A_3B_3A_4B_4\ket{d_1c_2d_3d_4}-A_1B_1A_2B_2\lvert B_3 \rvert^2\lvert B_4\rvert^2\ket{d_1d_2c_3c_4}\\&-A_1B_1A_2B_2\lvert B_3\rvert^2A_4B_4\ket{d_1d_2c_3d_4}-A_1B_1A_2B_2A_3B_3\lvert B_4\rvert^2\ket{d_1d_2d_3c_4}\\&-A_1B_1A_2B_2A_3B_3A_4B_4\ket{d_1d_2d_3d_4}].
        \label{e34}
    \end{split}
\end{equation}

Eq.\eqref{e34} has up to 16 terms; however, based on the correlation pointed out in the second set, only nine terms are needed to focus. For brevity, the state equation of this set only exhibits these states of interest as well. Accordingly, in this third set, there will be a total of nine conditions.

Regarding the contradiction with the LHV model, let's take an example of one of these nine states to see how this contradiction emerges. Considering the state $\ket{d_1d_2c_3c_4}$, its corresponding conditions in third set is $P(D_1D_2C_3C_4)>0$. It's not necessary to pull out all twelve conditions in the second set; conditions related to the first and second particles are enough. The LHV model can still be able to reproduce these conditions. For the third set, it is $D_1(\lambda)D_2(\lambda)C_3(\lambda)C_4(\lambda)=1$. For the second set, with the same local variable $\lambda$,

\begin{align}
    &D_1(\lambda)U_2(\lambda)V_3(\lambda)V_4(\lambda)=1,\label{e35}\\
    &D_1(\lambda)V_2(\lambda)V_3(\lambda)U_4(\lambda)=1,\label{e36}\\
    &D_1(\lambda)U_2(\lambda)V_3(\lambda)U_4(\lambda)=1,\label{e37}\\
    &U_1(\lambda)D_2(\lambda)V_3(\lambda)V_4(\lambda)=1,\label{e38}\\
    &V_1(\lambda)D_2(\lambda)U_3(\lambda)V_4(\lambda)=1,\label{e39}\\
    &U_1(\lambda)D_2(\lambda)U_3(\lambda)V_4(\lambda)=1.\label{e40}
\end{align}

From conditions \eqref{e35}$\rightarrow$\eqref{e40}, to reproduce the second set results, the LHV model had to cause $U_k=1 (k=1,2,3,4)$ on paired particles of the first and second particles, based on the correlation. As mentioned, with a single $D=1$, it can cause three possible outcomes. Consequently, there will be nine possible ways that the LHV model contradicts the first set. For instance, if choosing to follow the correlation \eqref{e35} and \eqref{e38}, it will be forced to cause $U_1(\lambda)U_2(\lambda)V_3(\lambda)V_4(\lambda)=1$, and the contradiction emerges. If choosing other correlations, this time, are \eqref{e35} and \eqref{e39}, then it is forced to cause $V_1(\lambda)U_2(\lambda)U_3(\lambda)V_4(\lambda)=1$, and the LHV model still contradicts the first set. So, with just a single state of interest in the third set, the number of ways to observe the Hardy's paradox is already considerable, and can be larger than the fully entangled system. The reasons for this increase, besides the number of states of interest in the third set, are also due to the diversity of correlations in the second set, as well as the large number of excluded states in the first set. Therefore, in terms of theoretical description, the cyclic entanglement offers a considerable number of ways to observe Hardy's nonlocality.

Regarding the $P_{\rm{success}}$, its value is calculated as the sum of the probability of the nine states exhibited in Eq.(\ref{e34}), which is
\begin{equation}
    \begin{split}
        P_{\text{success}}&=\langle\psi| C_1 C_2 D_3 D_4 |\psi\rangle+\langle\psi| C_1 D_2 D_3 C_4 |\psi\rangle+\langle\psi| C_1 D_2 D_3 D_4 |\psi\rangle\\&+\langle\psi| D_1 C_2 C_3 D_4 |\psi\rangle+\langle\psi| D_1 C_2 D_3 D_4 |\psi\rangle+\langle\psi| D_1 D_2 C_3 C_4 |\psi\rangle\\&+\langle\psi| D_1 D_2 C_3 D_4 |\psi\rangle+\langle\psi| D_1 D_2 D_3 C_4 |\psi\rangle+\langle\psi| D_1 D_2 D_3 D_4 |\psi\rangle.
        \label{e41}
    \end{split}
\end{equation}

\subsection{Constructing quantum circuits}

In this part, the way to construct four-qubit quantum circuits for each set of conditions will be discussed. Each quantum circuit is modified to match the purpose of each condition, and by simulating these circuits, the theoretical predictions from the previous parts will be verified. All quantum circuits used in the three sets of conditions are executed through IBM's QASM simulator

For the first set of conditions, the four qubits are initially prepared in the \{$\ket{c_k},\ket{d_k}$\} basis. The first mission is changing four qubit basis to \{$\ket{u_k},\ket{v_k}$\}. In the Mach-Zehnder experiment, this mission is carried out by the fixed beam splitters. In quantum circuits, the role of these beam splitters is played by rotation gates $R_y(\theta)$. After transforming the basis, four qubits are going through the post-selection process. As discussed, this process is achieved through the Toffoli gates, where two neighboring qubits form a pair to control an ancilla qubit. These Toffoli gates make the two qubits entangled and exclude specific states. The expected result after executing this circuit is that the measurement results of the nine vanished states in \eqref{e24} are all zero.

In the second set of conditions, the goal is to have states that indicate the correlations among qubits. Following the demonstration in the mathematical description, each qubit's basis is transformed from \{$\ket{u_k},\ket{v_k}$\} back to \{$\ket{c_k},\ket{d_k}$\}. To do this, the reverse rotation gate $R_y(-\theta)$ is applied to the chosen qubit. Similar to the theoretical description, this process is repeated four times, corresponding to changing the basis of each qubit one by one. There are in total of four circuits in this set. The expected result of each circuit is to obtain nonzero measurement probabilities of states, such as $\ket{d_1u_2v_3v_4}$, which show the correlation of a qubit with its paired qubits.

For the third set of conditions, all four qubits' basis are transformed back to \{$\ket{c_k},\ket{d_k}$\}. The method is similar to the second set. The rotation gates $R_y(-\theta)$ are utilized, but this time, these gates are implemented on all qubits instead of just one. There are two results expected to be obtained when executing this circuit. The first one is that the nine states of interest mentioned in Eq.(\ref{e34}) should have nonzero measurement probabilities. The second one is that the $P_{\rm{success}}$ has a nonzero value as well. 
Additionally, notable connections between theoretical descriptions and the quantum simulation results are $\ket{u_k},\ket{d_k}\equiv\ket{1}$, and $\ket{v_k},\ket{c_k}\equiv\ket{0}$. For instance, the state $\ket{u_1v_2v_3u_4}\equiv\ket{1001}$ or $\ket{c_1d_2d_3d_4}\equiv\ket{0111}$.

Furthermore, as discussed in previous parts, that $P_{\rm{success}}>0$ is a sign for the occurrence of Hardy's paradox, so it's natural to find the best setup so that $P_{\rm{success}}$ can be maximum. In accordance with Ref. \cite{PhysRevA.107.042210}, $P_{\rm{success}}$ depends on the value of $\theta$ used in the rotation gates $R_y$, $P_{\rm{success}}=f(\theta)$. In other words, finding the best setup is equivalent to finding the $\theta$ that can give maximum $P_{\rm{success}}$. The method is putting the circuit employed in the third set of conditions under a $\theta$ sweep test. In this test, the $\theta$ values range from 0 to $\pi$ radians in increment of $\frac{\pi}{18}$ radians. For each value of $\theta$, Eq.
\eqref{e41} is applied to compute the corresponding value of $P_{\rm{success}}$.

\section{Results}

Following the instructions in the previous section, quantum circuits corresponding to three sets of conditions were constructed. Before jumping into the main parts, there are four notices. The first notable thing is that the angle $\theta$ used in the quantum rotation gates $R_y(\theta)$ and $R_y(-\theta)$ is the same value. In this work, the chosen value of $\theta$ is $\theta\approx0.423\pi(rad)$.

The second notation is the way used to determine the theoretical values for each condition. Those values can be calculated by determining the three parameters: $A_k, B_k$, and $N$. Regarding the values of $A_k$ and $B_k$, they can be computed as $A_k=sin(\theta_k/2)$ and $B_k=\sqrt{1-A_k^2}$. Since all rotation gates in the circuits used the same $\theta$, then $A_1=A_2=A_3=A_4=A$ and $B_1=B_2=B_3=B_4=B$. Additionally, using $\theta\approx0.423\pi (rad)$, we obtained $A\approx0.617$ and $B\approx0.786$. Substituting these two results into Eq.(\ref{e31}), the value of $N$ is determined, which is approximately 1.26. Through substituting $A_k, B_k$, and $N$ into the state equations, the corresponding theoretical values are obtained. In the following parts, to verify the accuracy of the simulation results, their theoretical counterparts are provided. 

The next notation is about the depiction of quantum circuits. As discussed previously, there are four Toffoli gates in each circuit, and each of them controls a distinct ancilla qubit. Four distinct target qubits correspond to four separate lines, which causes space consumption. Therefore, to save space, the ancillas in all circuits in this article are placed in the same row.

The final notation is connections between theoretical descriptions and the quantum simulation results, which are represented through binary strings. Specifically, $\ket{u_k},\ket{d_k}\equiv\ket{1}$, and $\ket{v_k},\ket{c_k}\equiv\ket{0}$. For instance, the state $\ket{u_1v_2v_3u_4}\equiv\ket{1001}$ or $\ket{c_1d_2d_3d_4}\equiv\ket{0111}$.

\subsection{The first set of conditions}

\begin{figure}[H]
    \centering
    \includegraphics[width=1\textwidth,keepaspectratio]{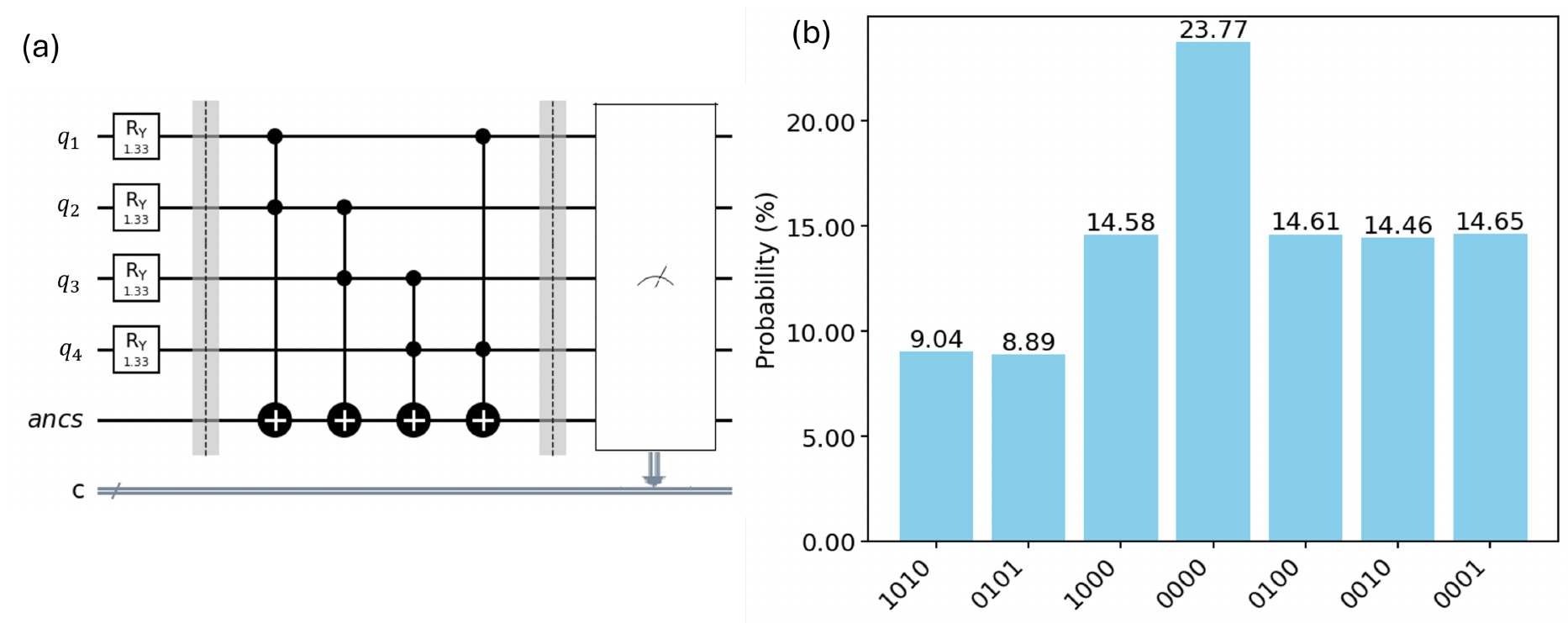}
    \vspace*{8pt}
    \caption{\justifying(a) Quantum circuit that is used to examine the first set of conditions. Four qubits are initially prepared in \{$\ket{c_k},\ket{d_k}$\}. By applying $Ry(\theta=0.423\pi)$ to each of them, they are all transformed to \{$\ket{u_k},\ket{v_k}$\}. Through Toffoli gates, they are made to be entangled with their neighbors, and the post-selection process is carried out. In particular, the nine states listed in \eqref{e24} are made to be undetectable, or have zero measurement probability. (b) Measurement result of states that can be detected from the circuit (2a).}
    \label{2}
\end{figure}

Beginning with the first set of conditions, the mission of the quantum circuit of this set is to exclude the nine vanished states \eqref{e24}. The quantum circuit is depicted in Fig. \ref{2}(a). In detail, four qubits are prepared in the ground state, which corresponds to $\ket{c_1}\otimes\ket{c_2}\otimes\ket{c_3}\otimes\ket{c_4}$. The rotation gates $R_y(\theta)$ are applied to each qubit to transform their basis from \{$\ket{c_k},\ket{d_k}$\} to \{$\ket{u_k},\ket{v_k}$\}. Subsequently, the qubits undergo Toffoli gates as described to become entangled, and for the post-selection process. 

Fig. \ref{2}(b) illustrates the measurement results of the circuit Fig. \ref{2}(a). It comprises seven states, matching the number of states described in Eq.\eqref{e30}. Moreover, from Fig. \ref{2}(a), nine states do not appear, or we can say that they have zero measurement probability; specifically, they are

\begin{equation}
    \{\ket{0011}, \ket{0110}, \ket{0111}, \ket{1001}, \ket{1011}, \ket{1100}, \ket{1101}, \ket{1110}, \ket{1111}\}.
        \label{e42}
\end{equation}

These states are all equivalent to the vanished states listed in \eqref{e24}. It indicates that the construction for the first circuit satisfies the target in the first set of conditions, causing the measurement probability of these states to become zero.

To verify the accuracy of each simulation result of the first set, the comparison between these results and their theoretical counterparts is shown in the following table

\begin{table}[h]
\caption{Comparison of simulation results and the theoretical calculations of the first set of conditions. 
The discrepancies between theory and simulation in this set are all below 0.7\%. \label{tab1}}
{\tabcolsep13pt
\begin{tabular}{|c|c|c|c|c|c|}
\hline
States & \multicolumn{2}{c|}{Probability (\%)} & States & \multicolumn{2}{c|}{Probability (\%)} \\
\cline{2-3} \cline{5-6}
& Simulation & Theory & & Simulation & Theory \\
\hline
0000 & 23.77 & 23.13 & 0101 & 8.89 & 8.78 \\
0001 & 14.65 & 14.25 & 1000 & 14.58 & 14.25 \\
0010 & 14.46 & 14.25 & 1010 & 9.04 & 8.78 \\
0100 & 14.61 & 14.25 & ---  & ---  & ---  \\
\hline
\end{tabular}}
\end{table}

As shown in Table \ref{tab1}, the theoretical results are nearly equal to their simulation calculations, with the largest deviation being only 0.64\%. Accordingly, based on Fig. \ref{2} and the comparison in Table \ref{tab1}, the simulation results of the first set are consistent with the theoretical descriptions. Specifically, the main target is achieved, where all vanished states have zero probability, and the deviations between the simulation and theoretical results are acceptable, with the maximum deviation being just 0.64\%. 

\subsection{The second set of conditions}

\begin{figure}[H]
        \centering
        \includegraphics[width=1\textwidth,keepaspectratio]{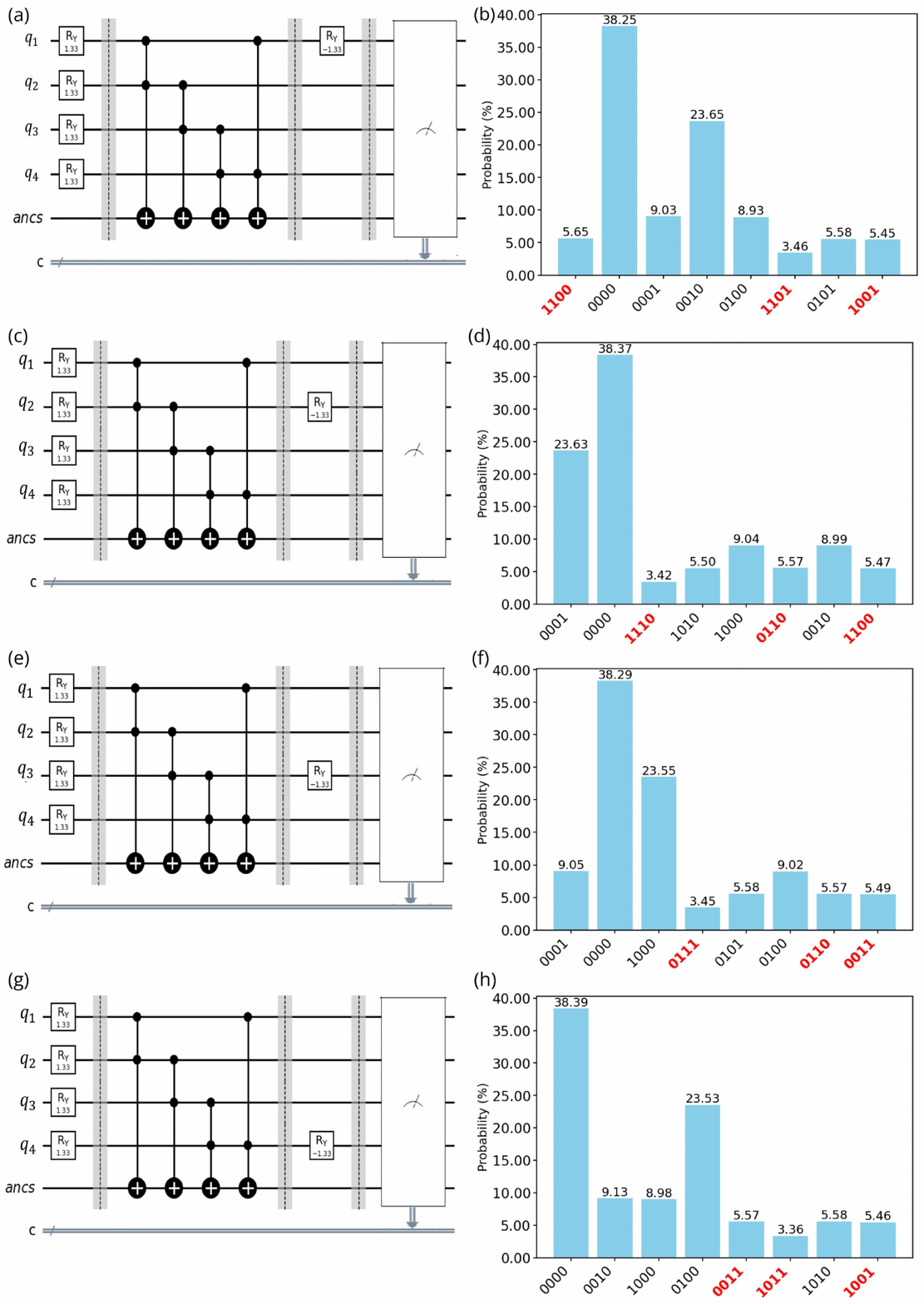}
        \vspace*{8pt}
        \caption{ Quantum circuits and the corresponding measurement results when transforming each qubit's basis. The states highlighted in red are states of interest. (a, b) The quantum circuit and its measurement result for the first condition of the second set. The first qubit's basis is transformed back to \{$\ket{c_k},\ket{d_k}$\} while the remaining qubits are retained in \{$\ket{u_k},\ket{v_k}$\}. From the circuit's results, we have three states of interest: $\ket{1001}, \ket{1101}, \ket{1100}$ corresponding to $\ket{d_1v_2v_3u_4}, \ket{d_1u_2v_3u_4}, \ket{d_1v_2v_3u_4}$. (c, d) The quantum circuit and its results when changing the second qubit's basis. The three states of interest in this condition are $\ket{1110}, \ket{1100}, \ket{0110}$. (e, f) The quantum circuit and its results when changing the third qubit's basis. The three states of interest are $\ket{0110}, \ket{0111}, \ket{0011}$. (g, h) The quantum circuit and its results when changing the fourth qubit's basis. The three states of interest are $\ket{1011}, \ket{0011}, \ket{1001}$.}
        \label{3}
    \end{figure}

Moving to the second set of conditions, the target is that the states implying the correlations, such as $\ket{d_1u_2v_3v_4}$, have nonzero results. As discussed, the basis of a qubit is changed back to \{$\ket{c_k},\ket{d_k}$\} while the others remain in \{$\ket{u_k},\ket{v_k}$\}. This set comprises four circuits, each corresponding to the basis transformation of one of four qubits. The $R_y(-\theta)$ gate is applied to a qubit that we want to change its basis. Fig.\ref{3} illustrates this set's quantum circuits and the corresponding results for these conditions.

Because the theoretical prediction when changing the basis of the first qubit was identified in Eq.\eqref{e33}, Fig. \ref{3}(a) and \ref{3}(b) will be chosen to analyze. Fig. \ref{3}(a) is the quantum circuit, in which the first qubit's basis is transformed back to \{$\ket{c_k},\ket{d_k}$\} while the remaining qubits are still measured in \{$\ket{u_k},\ket{v_k}$\}. Meanwhile, Fig. \ref{3}(b) shows the corresponding measurement result, which includes eight states with nonzero measurement probabilities, consistent with the number of states predicted in Eq.\eqref{e33}. Among these states, the states of interest, which correspond to $\ket{d_1v_2v_3u_4},\ket{d_1u_2v_3u_4},\ket{d_1u_2v_3v_4}$, are $\ket{1001}, \ket{1101}, \ket{1100}$ respectively. With the nonzero results of these three states, the theoretical assumption about the correlations among particles in the cyclic entanglement is confirmed. That is, a particle can only affect the results of its entangled partners. This pattern extends to other qubits as well. For instance, Fig. \ref{3}(d) shows the results when changing the second qubit basis. There are three highlighted states $\ket{1110}, \ket{1100}, \ket{0110}$ corresponding to $\ket{u_1d_2u_3v_4}, \ket{u_1d_2v_3v_4}, \ket{v_1d_2u_3v_4}$. These states imply that $D_2=1$ only causes $U_1=1$, $U_3=1$, or $U_1=U_3=1$, and as described, the second particle is entangled with the first and third particles.

The accuracy between the simulation and the theory of the second set can also be validated by the same method as in the first set. For instance, by substituting $A\approx0.617, B\approx0.786, N\approx1.26$ into Eq.\eqref{e33}, the theoretical results of the first condition (k=1) in the second set are achieved and shown in the Table \ref{tab2}.

\begin{table}[ph]
\caption{Comparison of simulation results and the theoretical calculations of the first condition of the second set. This table contains the theoretical calculations of Fig. 3b. The states of interest are still highlighted in red. The discrepancies between theory and simulation in this set are all below 0.8\%. \label{tab2}}
{\tabcolsep13pt
\begin{tabular}{|c|c|c|c|c|c|}
\hline
States & \multicolumn{2}{c|}{Probability (\%)} & States & \multicolumn{2}{c|}{Probability (\%)} \\
\cline{2-3} \cline{5-6}
& Simulation & Theory & & Simulation & Theory \\
\hline
0000 & 38.25 & 37.43 & 0101 & 5.58 & 5.43 \\
0001 & 9.03 & 8.80 & \textcolor{red}{1001} & 5.45 & 5.43 \\
0010 & 23.65 & 23.07 & \textcolor{red}{1100} & 5.65 & 5.42 \\
0100 & 8.93 & 8.80 & \textcolor{red}{1101} & 3.46 & 3.34 \\
\hline
\end{tabular}}
\end{table}

According to Table \ref{tab2}, the maximum discrepancy between simulation results and their theoretical calculations is about 0.8\%, which remains in the acceptable range. Accordingly, based on Fig. \ref{3} and Table \ref{tab2}, the simulation results of the second set are consistent with its theory. In particular, from Fig. \ref{3}, all states implying the correlation have nonzero probability, and the number of nonzero-probability states in each condition matches their theoretical description. Moreover, according to the Table \ref{tab2}, the deviations between the simulation and the theory are below 0.8\%.

\subsection{The third set of conditions}

\begin{figure}[H]
    \centering \includegraphics[width=1\textwidth,keepaspectratio]{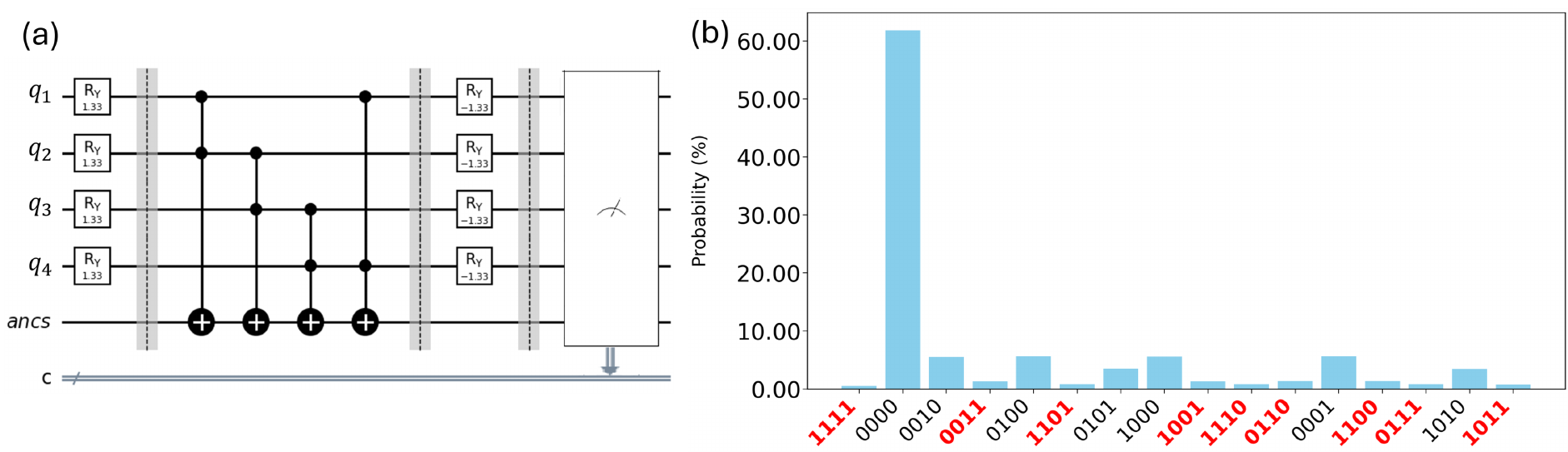}
    \vspace*{8pt}
    \caption{ (a,b) Quantum circuit and corresponding measurement's result for the third set of conditions. The states highlighted in red are states of interest. (a) All four qubits' basis are transformed back to \{$\ket{c_k},\ket{d_k}$\} by implementing the $R_y(-\theta)$ gates right after the Toffoli gates. (b) There are in total 16 states, matching the number of states described in Eq.\eqref{e34}.}
    \label{4}
\end{figure}

For the third set of conditions, all four qubits are transformed back to \{$\ket{c_k},\ket{d_k}$\} basis. The main goal of this set's circuit is that the predicted states of interest in Eq.\eqref{e34} have nonzero results. The circuit used in this set is illustrated in Fig. \ref{4}. 

Specifically, Fig. \ref{4}(a) is the circuit used to simulate this set, and its measurement results are shown in Fig. \ref{4}(b) and Table \ref{tab3}.

\begin{table}[H]
\caption{Specific measurement probabilities of 16 states shown in Figure (\ref{4}), and their theoretical calculations. The states of interest are highlighted in red. \label{tab3}}
{\tabcolsep13pt
\begin{tabular}{|c|c|c|c|c|c|}
\hline
States & \multicolumn{2}{c|}{Probability (\%)} & States & \multicolumn{2}{c|}{Probability (\%)} \\
\cline{2-3} \cline{5-6}
& Simulation & Theory & & Simulation & Theory \\
\hline
0000 & 61.83 & 61.62 & 1000 & 5.58 & 5.37 \\

0001 & 5.60 & 5.37 & \textcolor{red}{1001} & 1.30 & 1.27 \\

0010 & 5.51 & 5.37 & 1010 & 3.40 & 3.29 \\

\textcolor{red}{0011} & 1.30 & 1.27 & \textcolor{red}{1011} & 0.80 & 0.78 \\

0100 & 5.62 & 5.37 & \textcolor{red}{1100} & 1.33 & 1.27 \\

0101 & 3.45 & 3.29 & \textcolor{red}{1101} & 0.81 & 0.78 \\

\textcolor{red}{0110} & 1.31 & 1.27 & \textcolor{red}{1110} & 0.85 & 0.78 \\

\textcolor{red}{0111} & 0.80 & 0.78 & \textcolor{red}{1111} & 0.50 & 0.49 \\
\hline
\end{tabular}}
\end{table}

The nine states listed in \eqref{e43} correspond to the states of interest predicted in Eq.\eqref{e34}. The accuracy between the simulation results and their theory is shown in Table \ref{tab3}. The deviations are all below 0.4\%, and all states of interest have nonzero probability, indicating that the simulation results for this set are consistent with the theoretical predictions. 

\begin{equation}
    \{\ket{0011}, \ket{0110}, \ket{0111}, \ket{1001}, \ket{1011}, \ket{1100}, \ket{1101},\ket{1110}, \ket{1111}\}.
    \label{e43}
\end{equation}

Another target of the third set is determining the value of $P_{\rm{success}}$. It is determined as the sum of measurement probabilities of the nine states of interest of the third set, specifically
\begin{equation}
    \begin{split}
        P_{\text{success}}&=1.3+1.31+0.8+1.3+0.8+1.33+0.81+0.85+0.50\\&=9 \%.
        \label{e44}
    \end{split}
\end{equation}
To verify the accuracy of this result, it's necessary to compare it with the theoretical calculation defined in Eq. (\ref{e41}). Substituting $A\approx0.617, B\approx0.786,N\approx1.26$ into Eq.\eqref{e34}, the probability of each state will be obtained. The theoretical result of $P_{\rm{success}}$  can be calculated explicitly as
{\small
\begin{equation}
    \begin{split}
        P_{\text{success}}&=N^2(||B_1|^2|B_2|^2A_3B_3A_4B_4|^2+||B_1|^2A_2B_2A_3B_3|B_4|^2|^2+||B_1|^2A_2B_2A_3B_3A_4B_4|^2\\&+|A_1B_1|B_2|^2|B_3|^2A_4B_4|^2+|A_1B_1|B_2|^2A_3B_3A_4B_4|^2+|A_1B_1A_2B_2|B_3|^2|B_4|^2|^2\\&+|A_1B_1A_2B_2|B_3|^2A_4B_4|^2+|A_1B_1A_2B_2A_3B_3|B_4|^2|^2+|A_1B_1A_2B_2A_3B_3A_4B_4|^2)
        \\&\approx1.26^2(0.008+0.008+0.0049+0.008+0.0049+0.008+0.0049+0.0049+0.0031)\\&\approx0.0868=8.68\%.
        \label{e45}
    \end{split}
\end{equation}
}
The theoretical result Eq.\eqref{e45} is 0.32\% smaller than its simulation result shown in Eq.\eqref{e44}. This difference is sufficiently small and can be considered acceptable. Accordingly, the simulation results of the third set of conditions can be regarded as consistent with their theoretical predictions.

\begin{figure}[H]
    \centering
    \includegraphics[width=0.7\textwidth,keepaspectratio]{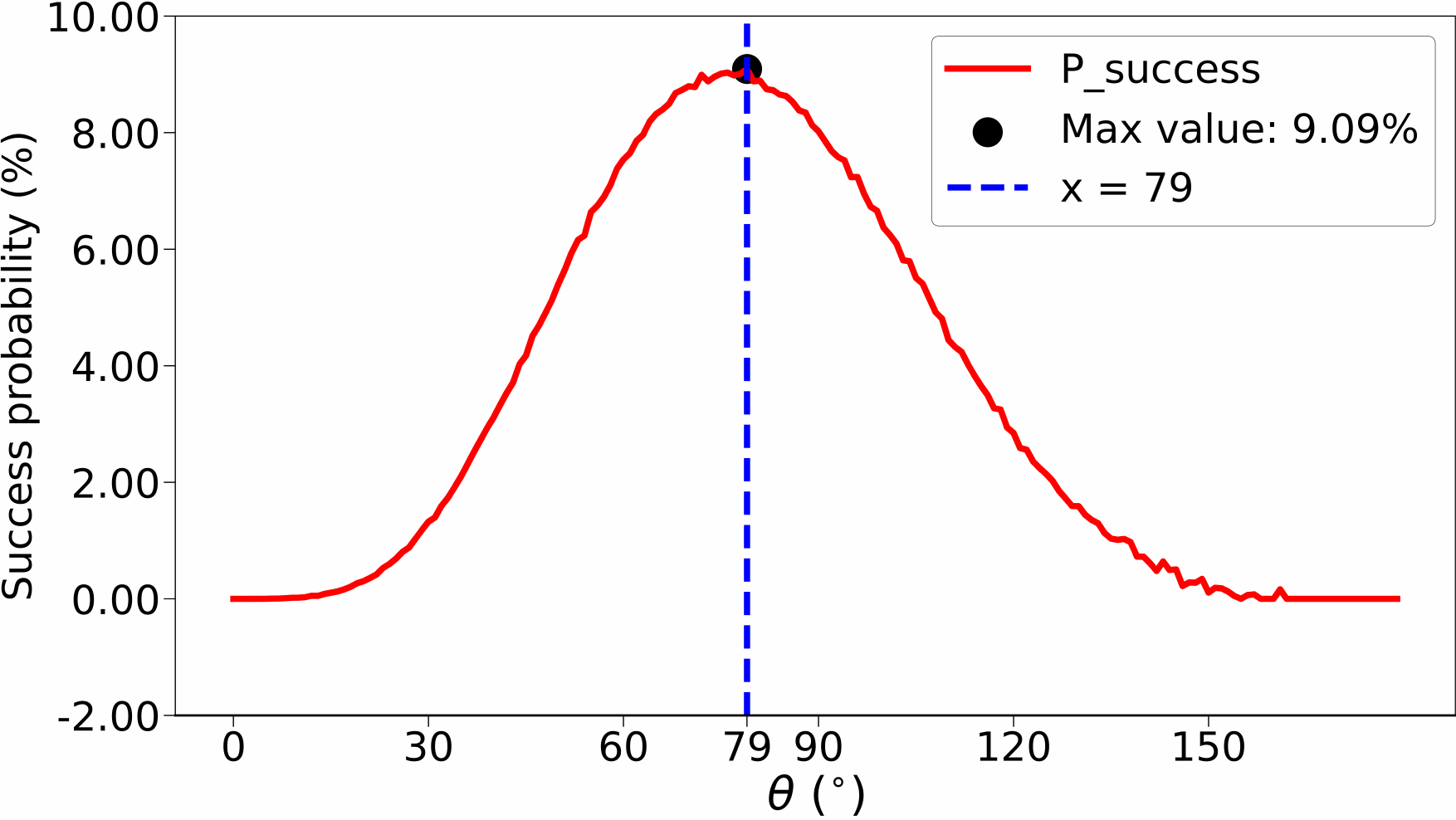}
    \vspace*{8pt}
    \caption{ $\theta$ sweep test. Using the circuit of the third set of conditions to examine. The range run from 0 to $\pi$ (rad) in increment of $\frac{\pi}{18}$ (rad). The maximum value of $P_{\text{success}}$ is approximately 9.09\% with the corresponding $\theta\approx0.439\pi (rad)\approx79(^{\circ})$.}
    \label{5}
\end{figure}

For the $\theta$ sweep test, Fig. \ref{5} shows the test result. The value of $\theta$ that was used in all the previous circuits is located in the vicinity of the value that gives the maximum value of $P_{\rm{success}}$. Specifically, after running the sweep test, the value of $\theta$ and the corresponding maximum value of $P_{\rm{success}}$ are: $\theta_{max}\approx0.439\pi (rad)\approx79 (^{\circ})$ and $P_{\rm{success}}\approx9.09\%$.

\section{Discussion}
\subsection{Limitations}
Although the approach used to determine the state equations can give a precise description of the corresponding quantum circuits, it's not an appropriate way to be used when the system size increases, especially when $n>4$. The main reason is that this approach is manual, so using this method may cost a lot of time in large systems. For instance, to determine the equation Eq.\eqref{e34} of the third set, all nine minus terms in Eq.\eqref{e29} have to changed to the \{$\ket{c_k},\ket{d_k}$\} basis. Moreover, each term comprises four members. Therefore, it will be time-consuming, and this approach might not be optimized for large systems. 

Our manual computation method also shows a drawback when demonstrating Hardy's paradox in systems with a large number of particles. Apparently, in larger systems, the correlation pattern remains consistent, which means particles only influence the results of those that are entangled with them. Therefore, the states of interest in each condition can still be determined by following this pattern, and Hardy's nonlocality as well. However, when the system's size increases, the number of states of interest in the third set increases as well. This number might increase considerably. Consequently, it will be challenging and time-consuming to compute the $P_{\rm{success}}$ by using the manual method applied in this work.

When conducting the $\theta$ sweep test, we observed that the peak's value was not fixed. The circuit was still Fig. 4a, and it was repeated 2048*200 times, or in short 2048*200 shots, the peak's values ran from $0.411\pi$ to $0.439\pi(rad)$, corresponding to the maximum value of $P_{\rm{success}}$ ran from 9.05\% to 9.22\%. Therefore, we believe that the true value of the peak $\theta$ may be located in the range $0.411\pi$ to $0.439\pi(rad)$ and might be achieved by increasing the number of shots. However, due to the technical limitations, we could not verify this hypothesis.

Additionally, we attempted to generalize this cyclic entanglement pattern to an n-particle system by examining similar circuit setups but with a larger number of qubits. Larger systems require longer circuit execution times and are more susceptible to errors. We reached our limit when executing an eight-qubit circuit with 1024*200 shots. In the case of $n=8$, we had to use a high number of shots when executing each circuit to mitigate errors and deviations as much as possible. However, the result's signal still exhibited considerable noise, and the execution time was too long.

\subsection{Results on a practical IBM's backend}
\begin{table}[ht]
\caption{IBM Brisbane's calibration data of the first six qubits, which are recorded at July 1st, 2025. The columns show the relaxation time $T_1$, decoherence time $T_2$, frequency, anharmonicity, and the readout length. \label{tab4}}
{\tabcolsep13pt
\begin{tabular}{|c|c|c|c|c|c|}
\hline
\textbf{Qubit} & \textbf{T1} & \textbf{T2} & \textbf{Freq} & \textbf{Anharmoni-} & \textbf{Readout} \\
& ($\mu$s) & ($\mu$s) & (GHz) & \textbf{-city} (GHz) & ns \\ \hline
0 & 255.3  & 58.4   & 4.722 & -0.31198 & 1300 \\ 
1 & 186.55 & 229.34 & 4.815 & -0.30974 & 1300 \\ 
2 & 282.35 & 92.86  & 4.610 & -0.30909 & 1300 \\ 
3 & 287.89 & 369.03 & 4.876 & -0.30969 & 1300 \\ 
4 & 205.65 & 247.42 & 4.818 & -0.31037 & 1300 \\ 
5 & 204.61 & 174.39 & 4.734 & -0.31145 & 1300 \\ 
\hline
\end{tabular}}
\end{table}

Moreover, we executed the circuits used in three sets of conditions on a practical backend of IBM, specifically IBM Brisbane. The data for Brisbane's first six qubits are shown in Table \ref {tab4}. Due to resource constraints and the limited time available for backend usage, each circuit was executed on IBM Brisbane with only 2048 shots, which is significantly fewer than the number of shots used in the simulations.

Table \ref{tab5} presents not only the measurement results of the first set of conditions, specifically circuits (2a), obtained from IBM Brisbane, but also the comparison of these results with the simulations. The important point is that the nine states listed in the \eqref{e24}, which theoretically should have zero measurement probabilities and were confirmed as such in simulation, now have nonzero probabilities. Most of these nine states show deviations of less than 6\% from the simulation, with the largest deviation belonging to $\ket{1100}$ with 13.03\%, $\ket{0110}$ follows with 8.95\%. The appearance of nonzero probabilities of these nine states in the first set of conditions not only goes against the supposed function of the Toffoli gates but also makes it meaningless, since we lost the conditions that contradict the LHV model's results. This contradiction is the heart of demonstrating the Hardy-type nonlocality. In addition, the other states exhibit significant differences with their simulation results, which range from approximately 0.5 to 13\%.

\begin{table}[H]
    \caption{Comparison of measurement results from IBM Brisbane and simulation of quantum circuits of the first set of conditions. The circuit used to examine is circuit (2a). From the results obtained from Brisbane, the nine vanished states \eqref{e24} have the nonzero measurement probabilities, which is against the purpose of the setup and may cause the contradiction with the third set not to occur. \label{tab5}}
    {\tabcolsep13pt
    \begin{tabular}{|c|c|c|c|c|c|}
    \hline
    \textbf{States} & \multicolumn{2}{c|}{\textbf{Probability (\%)}} & 
    \textbf{States} & \multicolumn{2}{c|}{\textbf{Probability (\%)}} \\
    \cline{2-3} \cline{5-6}
     & Simulation & Brisbane &  & Simulation & Brisbane \\
    \hline
    0000 & 23.77 & 8.79 & 1000 & 14.58 & 6.12 \\
    0001 & 14.65 & 2.98 & 1001 & 0 & 1.57 \\
    0010 & 14.46 & 12.56 & 1010 & 9.04 & 7.69 \\
    0011 & 0 & 3.45 & 1011 & 0 & 1.57 \\
    0100 & 14.61 & 15.23 & 1100 & 0 & 13.03 \\
    0101 & 8.89 & 5.65 & 1101 & 0 & 4.40 \\
    0110 & 0 & 8.95 & 1110 & 0 & 5.49 \\
    0111 & 0 & 1.26 & 1111 & 0 & 1.26 \\
    \hline
    \end{tabular}}
\end{table}

\begin{table}[H]
        \caption{Comparison of measurement results from IBM Brisbane and simulation of four conditions of the second set, in which each qubit's basis is changed back to \{$\ket{c_k},\ket{d_k}$\} sequentially. States of interest of each condition are highlighted in red. Their deviations from the simulation calculations are less than 5\%. \label{tab6}}
        {\tabcolsep8pt
        \begin{tabular}{|c|c|c|c|c|c|c|}
        \hline
        \textbf{$R_y(-\theta)$ on} & \textbf{States} & \multicolumn{2}{c|}{\textbf{Probability (\%)}} 
        & \textbf{States} & \multicolumn{2}{c|}{\textbf{Probability (\%)}} \\
        \cline{3-4}\cline{6-7}
        & & Simulation & Brisbane &  & Simulation & Brisbane \\
        \hline
        
        \multirow{4}{*}{$q_1$} 
         & 0000 & 38.25 & 7.25 & 1000 & 0 & 3.62 \\ 
         
         & 0001 & 9.03 & 5.80 & \textcolor{red}{1001} & 5.45 & 2.17 \\ 
          
         & 0010 & 23.65 & 6.52 & 1010 & 0 & 5.07 \\
          
         & 0011 & 0 & 3.62 & 1011 & 0 & 2.90 \\
        
         & 0100 & 8.93 & 14.49 & \textcolor{red}{1100} & 5.65 & 9.42 \\
         
         & 0101 & 5.58 & 10.14 & \textcolor{red}{1101} & 3.46 & 4.35 \\

         & 0110 & 0 & 9.42 & 1110 & 0 & 5.80 \\ 

         & 0111 & 0 & 4.35 & 1111 & 0 & 5.07 \\
        \hline\noalign{\hrule height 1pt}

        \multirow{4}{*}{$q_2$} 
         & 0000 & 38.37 & 5.64 & 1000 & 9.04 & 6.67 \\
         
         & 0001 & 23.63 & 6.67 & 1001 & 0 & 2.56 \\
         
         & 0010 & 8.99 & 4.62 & 1010 & 5.50 & 5.64 \\
         
         & 0011 & 0 & 5.13 & 1011 & 0 & 5.64 \\
         & 0100 & 0 & 9.23 & \textcolor{red}{1100} & 5.47 & 9.74 \\
         & 0101 & 0 & 10.26 & 1101 & 0 & 4.62 \\
         & \textcolor{red}{0110} & 5.57 & 8.72 & \textcolor{red}{1110} & 3.42 & 3.08 \\
         & 0111 & 0 & 5.64 & 1111 & 0 & 6.15 \\
        \hline\noalign{\hrule height 1pt}

        \multirow{4}{*}{$q_3$} 
         & 0000 & 38.29 & 8.94 & 1000 & 23.55 & 4.34 \\
         
         & 0001 & 9.05 & 13.01 & 1001 & 0 & 3.15 \\
         
         & 0010 & 0 & 6.57 & 1010 & 0 & 4.99 \\
         
         & \textcolor{red}{0011} & 5.49 & 5.39 & 1011 & 0 & 3.15 \\
         & 0100 & 9.02 & 8.02 & 1100 & 0 & 8.67 \\
         & 0101 & 5.58 & 9.99  & 1101 & 0 & 6.18 \\
         & \textcolor{red}{0110} & 5.57 & 7.10 & 1110 & 0 & 3.29 \\
         & \textcolor{red}{0111} & 3.45 & 5.12 & 1111 & 0 & 2.10 \\
        \hline\noalign{\hrule height 1pt}
        
        \end{tabular}}
    \end{table}

\addtocounter{table}{-1}
\begin{table}[H]
        \caption{(Continue). \label{tab6x}}
        {\tabcolsep8pt
        \begin{tabular}{|c|c|c|c|c|c|c|}
        \hline
        \textbf{$R_y(-\theta)$ on} & \textbf{States} & \multicolumn{2}{c|}{\textbf{Probability (\%)}} 
        & \textbf{States} & \multicolumn{2}{c|}{\textbf{Probability (\%)}} \\
        \cline{3-4}\cline{6-7}
        & & Simulation & Brisbane &  & Simulation & Brisbane \\
        \hline
        \multirow{4}{*}{$q_4$} 
         & 0000 & 38.39 & 5.82 & 1000 & 8.98 & 6.93 \\
         
         & 0001 & 0 & 9.70 & \textcolor{red}{1001} & 5.46 & 8.03 \\
         
         & 0010 & 9.13 & 5.82 & 1010 & 5.58 & 7.76 \\
         
         & \textcolor{red}{0011} & 5.57 & 6.93 & \textcolor{red}{1011} & 3.36 & 7.48 \\
         & 0100 & 23.53 & 3.60 & 1100 & 0 & 4.71 \\
         & 0101 & 0 & 5.82 & 1101 & 0 & 5.82 \\
         & 0110 & 0 & 4.71 & 1110 & 0 & 7.48 \\
         & 0111 & 0 & 3.32 & 1111 & 0 & 6.09 \\
        \hline\noalign{\hrule height 1pt}
    \end{tabular}}
\end{table}

The results of the second set of conditions on the IBM Brisbane, and their deviations from the simulations, are shown in Table \ref{tab6}. There are a total of four circuits in this set, which are labeled from $q_1$ to $q_4$ as shown in the table.

States of interest in each circuit are still achieved from Brisbane's results, and their deviations are mostly within 5\%, which is acceptable. However, similar to the first set of conditions, the discrepancy with the simulation results is most evident in the difference in the number of states having nonzero probabilities. For instance, in Table \ref{tab4}, in the first circuit of the second set, in which the first qubit's basis is transformed back to \{$\ket{c_k},\ket{d_k}$\}, all 16 states have nonzero probabilities in the Brisbane's results, whereas the simulation identifies only eight such states, which is only half the number of results obtained from Brisbane. Furthermore, several states have significant deviations in their measurement probabilities compared to the simulation result. This pattern extends to other conditions in this set as well.

\begin{table}[H]
    \caption{Comparison of measurement results from IBM Brisbane and simulation of quantum circuits of the third set of conditions, in which all qubits' bases are transformed back to \{$\ket{c_k},\ket{d_k}$\}. The circuit used to examine is circuit (4a). The nine states of interest are highlighted in red, and their deviations from the simulation are less than 10\%. \label{tab7}}
    {\tabcolsep13pt
    \begin{tabular}{|c|c|c|c|c|c|}
    \hline
    \textbf{States} & \multicolumn{2}{c|}{\textbf{Probability (\%)}} & 
    \textbf{States} & \multicolumn{2}{c|}{\textbf{Probability (\%)}} \\
    \cline{2-3} \cline{5-6}
     & Simulation & Brisbane &  & Simulation & Brisbane \\
    \hline
    0000 & 61.83 & 8.64 & 1000 & 5.58 & 5.00 \\
    0001 & 5.60 & 6.82 & \textcolor{red}{1001} & 1.30 & 1.82 \\
    0010 & 5.51 & 10.91 & 1010 & 3.40 & 5.00 \\
    \textcolor{red}{0011} & 1.30 & 6.82 & \textcolor{red}{1011} & 1.30 & 1.82 \\
    0100 & 5.62 & 8.64 & \textcolor{red}{1100} & 1.33 & 9.55 \\
    0101 & 3.45 & 8.18 & \textcolor{red}{1101} & 0.81 & 4.09 \\
    \textcolor{red}{0110} & 1.31 & 8.18 & \textcolor{red}{1110} & 0.85 & 4.09 \\
    \textcolor{red}{0111} & 0.80 & 7.27 & \textcolor{red}{1111} & 0.50 & 4.09 \\
    \hline
    \end{tabular}}
\end{table}

Table \ref{tab7} is the comparison of the simulation and Brisbane results of the third set of conditions. Unlike the first and second sets, the number of nonzero probability states of this set in the simulation and in IBM Brisbane is equal, which is sixteen states. The discrepancy with the simulation results only lies in the deviations of measurement probabilities. The nine states of interest have the deviations of less than 10\% compared to the simulation results. However, as previously mentioned, when analysing the deviations of the first set, these nonzero probabilities are no longer valuable for demonstrating Hardy-type nonlocality, since in the IBM Brisbane results, these vanished states were not excluded to establish the later contradiction. Additionally, the remaining states also have significant deviations from their simulation calculations. 

In summary, the results obtained from IBM Brisbane for the three sets of conditions showed a considerable difference compared to their simulation calculations. Specifically, the discrepancies lie in two aspects: the difference in the number of states having nonzero probabilities and the deviations in measurement probabilities. The difference in the number of nonzero probability states is evident in the first and third sets. Meanwhile, a notable example of the deviation in measurement probability between the Brisbane result and the simulation is the state $\ket{0000}$, with the deviation reaching up to 30 and 50\% in the second and third sets.

We were not able to give precise answers to what caused significant deviations between Brisbane results and the simulation calculations. However, in an effort to locate the reasons, we conducted several simple experiments, and their results helped us to narrow down the potential sources of these discrepancies. These experiments, detailed later in the Appendix, provided valuable insights that informed our hypotheses about the underlying errors. We propose two reasons contributing to these discrepancies: the number of control qubits used in a Toffoli gate and the value of $\theta$ used in $Ry$ gates. Based on our observations, the number of qubits used in a Toffoli gate may be the major source of error. Specifically, if this number is larger than one, it will start to cause significant errors. Furthermore, if a qubit is used as a control qubit in more than one gate, it may exacerbate the error. The value of $\theta$ may serve as the secondary factor that makes the error worse. In particular, if $\theta$ is increased, the error will also start to occur, but it will not be as significant compared to the error caused by multiple control qubits. Additionally, the fact that each circuit was run for only 1024*2 shots on Brisbane may also give us results that are not sufficiently good.

\subsection{Compare to the fully entangled system using the same approach}
Although this work builds upon the approach of Ref. \cite{PhysRevA.107.042210} to observe Hardy-type nonlocality, there is a difference between the two studies. The most significant difference between our work and Ref. \cite{PhysRevA.107.042210} lies in the entanglement configuration, which was discussed in detail in the first section.

A further difference is the value of $P_{\rm{success}}$. In both works, there are multiple states of interest in the third set. However, for a four-particle system, the value of $P_{\rm{success}}$ is reported approximately 12.5\% in Ref. \cite{PhysRevA.107.042210}; whereas in the cyclic entanglement considered here, the corresponding value is approximately 9\%. This distinction can be traced back to the different criteria used to select the states of interest. 

In the fully entangled system, the selection rule requires only that at least two particles yield D=1, which allows a wider range of contributing states. Meanwhile, in the cyclic entanglement, the criterion is more restrictive: at least two particles belonging to the same entangled pair must have D=1. As a result, certain states, such as $\ket{d_1c_2d_3c_4}$ and $\ket{c_1d_2c_3d_4}$, contribute to $P_{\rm{success}}$ in the fully entangled cases but are dismissed in the cyclic configuration. 

For instance, consider the two states $\ket{d_1c_2d_3c_4}$ and $\ket{c_1d_2c_3d_4}$. In the fully entangled setups, if the LHV model attempts to reproduce $D_1(\lambda)C_2(\lambda)D_3(\lambda)C_4(\lambda)=1$ or $C_1(\lambda)D_2(\lambda)C_3(\lambda)D_4(\lambda)=1$, it will still end up leads to $U_1(\lambda)U_2(\lambda)U_3(\lambda)U_4(\lambda)=1$, which contradicts with the first set. However, in the cyclic configuration, the LHV model will just be forced to cause some states, such as $V_1(\lambda)U_2(\lambda)V_3(\lambda)U_4(\lambda)=1$ or $V_1(\lambda)U_2(\lambda)V_3(\lambda)V_4(\lambda)=1$, which are not excluded states in the first set, and nothing is contradicted.

Consequently, although the cyclic entanglement admits a larger variety of conditions and various ways to form the paradox, the more restrictive selection rule for states of interest leads to a smaller $P_{\rm{success}}$ compared to the fully entangled systems.

\section{Conclusions}

In conclusion, the Hardy-type nonlocality in a cyclic entanglement configuration can still be observed as normally as in the fully entangled systems. However, cyclic entanglement provides more ways to construct the contradiction with the LHV model than in the fully entangled system. In particular, due to multiple excluded states in the first set, various correlations in the second set, and multiple states of interest in the third set as well, the number of ways and conditions used to observe Hardy's paradox increases considerably. However, because of a more restrictive selection criterion for the states of interest, $P_{\rm{success}}$ is smaller than that in the fully entangled cases. Additionally, the correlation pattern in the cyclic entanglement is also pointed out, which is that a particle can only affect its entangled partners' results. 

The simulation results and the theoretical counterparts show high consistency. In particular, the deviations of simulation results from the theory of all three sets are below 1\%. With $\theta\approx0.423\pi(rad)$,  the value of $P_{\rm{success}}$ is found to be 9\%. This value's discrepancy with its theory is also just 0.32\%. Additionally, by performing the $\theta$ sweep test, the $\theta$ value that can give the maximum value of $P_{\rm{success}}$ has been determined, specifically, when $\theta\approx0.439\pi(rad)$ it can give us $P_{\rm{success}}\approx9.09\%$.
The consistency between simulation and theory shows that simulating quantum circuits is a useful tool to realize and confirm the quantum foundation experiments.

However, there are limitations emerged in our work. The approach used to determine state equations is manual. Therefore, using this approach will take a lot of time to describe the system when n increases. Additionally, if performing the sweep test repeatedly, the peak's location is not fixed; under our examination, the $\theta_{max}$ fluctuates in a range $[0.411\pi,0.439\pi](rad)$ corresponding to the fluctuation of $P_{\rm{success}}$ in a range [9.05\%,9.22\%]. The last thing is that we also attempted to execute the circuits of the three sets on a practical quantum backend, specifically IBM Brisbane. Nevertheless, the results show significant deviations from their simulation calculations. The discrepancies can be up to approximately 30-50\%.

\section*{Acknowledgments}
This work was supported by Vietnam National University Hanoi under Grant Number QG.24.105. The authors declare no conflicts of interest regarding this manuscript.


\bibliographystyle{apsrev4-1}
\bibliography{bib}

\begin{thebibliography}{39}%
\makeatletter
\providecommand \@ifxundefined [1]{%
 \@ifx{#1\undefined}
}%
\providecommand \@ifnum [1]{%
 \ifnum #1\expandafter \@firstoftwo
 \else \expandafter \@secondoftwo
 \fi
}%
\providecommand \@ifx [1]{%
 \ifx #1\expandafter \@firstoftwo
 \else \expandafter \@secondoftwo
 \fi
}%
\providecommand \natexlab [1]{#1}%
\providecommand \enquote  [1]{``#1''}%
\providecommand \bibnamefont  [1]{#1}%
\providecommand \bibfnamefont [1]{#1}%
\providecommand \citenamefont [1]{#1}%
\providecommand \href@noop [0]{\@secondoftwo}%
\providecommand \href [0]{\begingroup \@sanitize@url \@href}%
\providecommand \@href[1]{\@@startlink{#1}\@@href}%
\providecommand \@@href[1]{\endgroup#1\@@endlink}%
\providecommand \@sanitize@url [0]{\catcode `\\12\catcode `\$12\catcode
  `\&12\catcode `\#12\catcode `\^12\catcode `\_12\catcode `\%12\relax}%
\providecommand \@@startlink[1]{}%
\providecommand \@@endlink[0]{}%
\providecommand \url  [0]{\begingroup\@sanitize@url \@url }%
\providecommand \@url [1]{\endgroup\@href {#1}{\urlprefix }}%
\providecommand \urlprefix  [0]{URL }%
\providecommand \Eprint [0]{\href }%
\providecommand \doibase [0]{http://dx.doi.org/}%
\providecommand \selectlanguage [0]{\@gobble}%
\providecommand \bibinfo  [0]{\@secondoftwo}%
\providecommand \bibfield  [0]{\@secondoftwo}%
\providecommand \translation [1]{[#1]}%
\providecommand \BibitemOpen [0]{}%
\providecommand \bibitemStop [0]{}%
\providecommand \bibitemNoStop [0]{.\EOS\space}%
\providecommand \EOS [0]{\spacefactor3000\relax}%
\providecommand \BibitemShut  [1]{\csname bibitem#1\endcsname}%
\let\auto@bib@innerbib\@empty
\bibitem [{\citenamefont {Einstein}\ \emph {et~al.}(1935)\citenamefont
  {Einstein}, \citenamefont {Podolsky},\ and\ \citenamefont
  {Rosen}}]{einstein1935can}%
  \BibitemOpen
  \bibfield  {author} {\bibinfo {author} {\bibfnamefont {A.}~\bibnamefont
  {Einstein}}, \bibinfo {author} {\bibfnamefont {B.}~\bibnamefont {Podolsky}},
  \ and\ \bibinfo {author} {\bibfnamefont {N.}~\bibnamefont {Rosen}},\
  }\href@noop {} {\bibfield  {journal} {\bibinfo  {journal} {Physical review}\
  }\textbf {\bibinfo {volume} {47}},\ \bibinfo {pages} {777} (\bibinfo {year}
  {1935})}\BibitemShut {NoStop}%
\bibitem [{\citenamefont {Bell}(1964)}]{bell1964einstein}%
  \BibitemOpen
  \bibfield  {author} {\bibinfo {author} {\bibfnamefont {J.~S.}\ \bibnamefont
  {Bell}},\ }\href@noop {} {\bibfield  {journal} {\bibinfo  {journal} {Physics
  Physique Fizika}\ }\textbf {\bibinfo {volume} {1}},\ \bibinfo {pages} {195}
  (\bibinfo {year} {1964})}\BibitemShut {NoStop}%
\bibitem [{\citenamefont {Clauser}\ \emph {et~al.}(1969)\citenamefont
  {Clauser}, \citenamefont {Horne}, \citenamefont {Shimony},\ and\
  \citenamefont {Holt}}]{clauser1969proposed}%
  \BibitemOpen
  \bibfield  {author} {\bibinfo {author} {\bibfnamefont {J.~F.}\ \bibnamefont
  {Clauser}}, \bibinfo {author} {\bibfnamefont {M.~A.}\ \bibnamefont {Horne}},
  \bibinfo {author} {\bibfnamefont {A.}~\bibnamefont {Shimony}}, \ and\
  \bibinfo {author} {\bibfnamefont {R.~A.}\ \bibnamefont {Holt}},\ }\href@noop
  {} {\bibfield  {journal} {\bibinfo  {journal} {Physical review letters}\
  }\textbf {\bibinfo {volume} {23}},\ \bibinfo {pages} {880} (\bibinfo {year}
  {1969})}\BibitemShut {NoStop}%
\bibitem [{\citenamefont {Hardy}(1992)}]{PhysRevLett.68.2981}%
  \BibitemOpen
  \bibfield  {author} {\bibinfo {author} {\bibfnamefont {L.}~\bibnamefont
  {Hardy}},\ }\href {\doibase 10.1103/PhysRevLett.68.2981} {\bibfield
  {journal} {\bibinfo  {journal} {Phys. Rev. Lett.}\ }\textbf {\bibinfo
  {volume} {68}},\ \bibinfo {pages} {2981} (\bibinfo {year}
  {1992})}\BibitemShut {NoStop}%
\bibitem [{\citenamefont {Hardy}(1993)}]{PhysRevLett.71.1665}%
  \BibitemOpen
  \bibfield  {author} {\bibinfo {author} {\bibfnamefont {L.}~\bibnamefont
  {Hardy}},\ }\href {\doibase 10.1103/PhysRevLett.71.1665} {\bibfield
  {journal} {\bibinfo  {journal} {Phys. Rev. Lett.}\ }\textbf {\bibinfo
  {volume} {71}},\ \bibinfo {pages} {1665} (\bibinfo {year}
  {1993})}\BibitemShut {NoStop}%
\bibitem [{\citenamefont {Tran}\ \emph {et~al.}(2023)\citenamefont {Tran},
  \citenamefont {Nguyen}, \citenamefont {Ho},\ and\ \citenamefont
  {Nguyen}}]{PhysRevA.107.042210}%
  \BibitemOpen
  \bibfield  {author} {\bibinfo {author} {\bibfnamefont {D.~M.}\ \bibnamefont
  {Tran}}, \bibinfo {author} {\bibfnamefont {V.-D.}\ \bibnamefont {Nguyen}},
  \bibinfo {author} {\bibfnamefont {L.~B.}\ \bibnamefont {Ho}}, \ and\ \bibinfo
  {author} {\bibfnamefont {H.~Q.}\ \bibnamefont {Nguyen}},\ }\href {\doibase
  10.1103/PhysRevA.107.042210} {\bibfield  {journal} {\bibinfo  {journal}
  {Phys. Rev. A}\ }\textbf {\bibinfo {volume} {107}},\ \bibinfo {pages}
  {042210} (\bibinfo {year} {2023})}\BibitemShut {NoStop}%
\bibitem [{\citenamefont {Jordan}(1994)}]{PhysRevA.50.62}%
  \BibitemOpen
  \bibfield  {author} {\bibinfo {author} {\bibfnamefont {T.~F.}\ \bibnamefont
  {Jordan}},\ }\href {\doibase 10.1103/PhysRevA.50.62} {\bibfield  {journal}
  {\bibinfo  {journal} {Phys. Rev. A}\ }\textbf {\bibinfo {volume} {50}},\
  \bibinfo {pages} {62} (\bibinfo {year} {1994})}\BibitemShut {NoStop}%
\bibitem [{\citenamefont {Kar}(1997)}]{PhysRevA.56.1023}%
  \BibitemOpen
  \bibfield  {author} {\bibinfo {author} {\bibfnamefont {G.}~\bibnamefont
  {Kar}},\ }\href {\doibase 10.1103/PhysRevA.56.1023} {\bibfield  {journal}
  {\bibinfo  {journal} {Phys. Rev. A}\ }\textbf {\bibinfo {volume} {56}},\
  \bibinfo {pages} {1023} (\bibinfo {year} {1997})}\BibitemShut {NoStop}%
\bibitem [{\citenamefont {hua Wu}\ and\ \citenamefont {hua
  Xie}(1996)}]{WU1996129}%
  \BibitemOpen
  \bibfield  {author} {\bibinfo {author} {\bibfnamefont {X.}~\bibnamefont {hua
  Wu}}\ and\ \bibinfo {author} {\bibfnamefont {R.}~\bibnamefont {hua Xie}},\
  }\href {\doibase https://doi.org/10.1016/0375-9601(95)00957-4} {\bibfield
  {journal} {\bibinfo  {journal} {Physics Letters A}\ }\textbf {\bibinfo
  {volume} {211}},\ \bibinfo {pages} {129} (\bibinfo {year}
  {1996})}\BibitemShut {NoStop}%
\bibitem [{\citenamefont {Ghosh}\ \emph {et~al.}(1998)\citenamefont {Ghosh},
  \citenamefont {Kar},\ and\ \citenamefont {Sarkar}}]{GHOSH1998249}%
  \BibitemOpen
  \bibfield  {author} {\bibinfo {author} {\bibfnamefont {S.}~\bibnamefont
  {Ghosh}}, \bibinfo {author} {\bibfnamefont {G.}~\bibnamefont {Kar}}, \ and\
  \bibinfo {author} {\bibfnamefont {D.}~\bibnamefont {Sarkar}},\ }\href
  {\doibase https://doi.org/10.1016/S0375-9601(98)00306-5} {\bibfield
  {journal} {\bibinfo  {journal} {Physics Letters A}\ }\textbf {\bibinfo
  {volume} {243}},\ \bibinfo {pages} {249} (\bibinfo {year}
  {1998})}\BibitemShut {NoStop}%
\bibitem [{\citenamefont {Wu}\ \emph {et~al.}(2000)\citenamefont {Wu},
  \citenamefont {Zong},\ and\ \citenamefont {Pang}}]{WU2000221}%
  \BibitemOpen
  \bibfield  {author} {\bibinfo {author} {\bibfnamefont {X.-H.}\ \bibnamefont
  {Wu}}, \bibinfo {author} {\bibfnamefont {H.-S.}\ \bibnamefont {Zong}}, \ and\
  \bibinfo {author} {\bibfnamefont {H.-R.}\ \bibnamefont {Pang}},\ }\href
  {\doibase https://doi.org/10.1016/S0375-9601(00)00667-8} {\bibfield
  {journal} {\bibinfo  {journal} {Physics Letters A}\ }\textbf {\bibinfo
  {volume} {276}},\ \bibinfo {pages} {221} (\bibinfo {year}
  {2000})}\BibitemShut {NoStop}%
\bibitem [{\citenamefont {Cereceda}(2004)}]{CERECEDA2004433}%
  \BibitemOpen
  \bibfield  {author} {\bibinfo {author} {\bibfnamefont {J.~L.}\ \bibnamefont
  {Cereceda}},\ }\href {\doibase
  https://doi.org/10.1016/j.physleta.2004.06.004} {\bibfield  {journal}
  {\bibinfo  {journal} {Physics Letters A}\ }\textbf {\bibinfo {volume}
  {327}},\ \bibinfo {pages} {433} (\bibinfo {year} {2004})}\BibitemShut
  {NoStop}%
\bibitem [{\citenamefont {Jiang}\ \emph {et~al.}(2018)\citenamefont {Jiang},
  \citenamefont {Xu}, \citenamefont {Su}, \citenamefont {Pati},\ and\
  \citenamefont {Chen}}]{PhysRevLett.120.050403}%
  \BibitemOpen
  \bibfield  {author} {\bibinfo {author} {\bibfnamefont {S.-H.}\ \bibnamefont
  {Jiang}}, \bibinfo {author} {\bibfnamefont {Z.-P.}\ \bibnamefont {Xu}},
  \bibinfo {author} {\bibfnamefont {H.-Y.}\ \bibnamefont {Su}}, \bibinfo
  {author} {\bibfnamefont {A.~K.}\ \bibnamefont {Pati}}, \ and\ \bibinfo
  {author} {\bibfnamefont {J.-L.}\ \bibnamefont {Chen}},\ }\href {\doibase
  10.1103/PhysRevLett.120.050403} {\bibfield  {journal} {\bibinfo  {journal}
  {Phys. Rev. Lett.}\ }\textbf {\bibinfo {volume} {120}},\ \bibinfo {pages}
  {050403} (\bibinfo {year} {2018})}\BibitemShut {NoStop}%
\bibitem [{\citenamefont {Cabello}\ \emph {et~al.}(2008)\citenamefont
  {Cabello}, \citenamefont {G{\"u}hne}, \citenamefont {Moreno},\ and\
  \citenamefont {Rodr{\'\i}guez}}]{cabello2008nonlocality}%
  \BibitemOpen
  \bibfield  {author} {\bibinfo {author} {\bibfnamefont {A.}~\bibnamefont
  {Cabello}}, \bibinfo {author} {\bibfnamefont {O.}~\bibnamefont {G{\"u}hne}},
  \bibinfo {author} {\bibfnamefont {P.}~\bibnamefont {Moreno}}, \ and\ \bibinfo
  {author} {\bibfnamefont {D.}~\bibnamefont {Rodr{\'\i}guez}},\ }\href@noop {}
  {\bibfield  {journal} {\bibinfo  {journal} {Laser physics}\ }\textbf
  {\bibinfo {volume} {18}},\ \bibinfo {pages} {335} (\bibinfo {year}
  {2008})}\BibitemShut {NoStop}%
\bibitem [{\citenamefont {Gachechiladze}\ \emph {et~al.}(2016)\citenamefont
  {Gachechiladze}, \citenamefont {Budroni},\ and\ \citenamefont
  {G\"uhne}}]{PhysRevLett.116.070401}%
  \BibitemOpen
  \bibfield  {author} {\bibinfo {author} {\bibfnamefont {M.}~\bibnamefont
  {Gachechiladze}}, \bibinfo {author} {\bibfnamefont {C.}~\bibnamefont
  {Budroni}}, \ and\ \bibinfo {author} {\bibfnamefont {O.}~\bibnamefont
  {G\"uhne}},\ }\href {\doibase 10.1103/PhysRevLett.116.070401} {\bibfield
  {journal} {\bibinfo  {journal} {Phys. Rev. Lett.}\ }\textbf {\bibinfo
  {volume} {116}},\ \bibinfo {pages} {070401} (\bibinfo {year}
  {2016})}\BibitemShut {NoStop}%
\bibitem [{\citenamefont {Barnea}\ \emph {et~al.}(2015)\citenamefont {Barnea},
  \citenamefont {P\"utz}, \citenamefont {Brask}, \citenamefont {Brunner},
  \citenamefont {Gisin},\ and\ \citenamefont {Liang}}]{PhysRevA.91.032108}%
  \BibitemOpen
  \bibfield  {author} {\bibinfo {author} {\bibfnamefont {T.~J.}\ \bibnamefont
  {Barnea}}, \bibinfo {author} {\bibfnamefont {G.}~\bibnamefont {P\"utz}},
  \bibinfo {author} {\bibfnamefont {J.~B.}\ \bibnamefont {Brask}}, \bibinfo
  {author} {\bibfnamefont {N.}~\bibnamefont {Brunner}}, \bibinfo {author}
  {\bibfnamefont {N.}~\bibnamefont {Gisin}}, \ and\ \bibinfo {author}
  {\bibfnamefont {Y.-C.}\ \bibnamefont {Liang}},\ }\href {\doibase
  10.1103/PhysRevA.91.032108} {\bibfield  {journal} {\bibinfo  {journal} {Phys.
  Rev. A}\ }\textbf {\bibinfo {volume} {91}},\ \bibinfo {pages} {032108}
  (\bibinfo {year} {2015})}\BibitemShut {NoStop}%
\bibitem [{\citenamefont {Wang}\ and\ \citenamefont
  {Markham}(2012)}]{PhysRevLett.108.210407}%
  \BibitemOpen
  \bibfield  {author} {\bibinfo {author} {\bibfnamefont {Z.}~\bibnamefont
  {Wang}}\ and\ \bibinfo {author} {\bibfnamefont {D.}~\bibnamefont {Markham}},\
  }\href {\doibase 10.1103/PhysRevLett.108.210407} {\bibfield  {journal}
  {\bibinfo  {journal} {Phys. Rev. Lett.}\ }\textbf {\bibinfo {volume} {108}},\
  \bibinfo {pages} {210407} (\bibinfo {year} {2012})}\BibitemShut {NoStop}%
\bibitem [{\citenamefont {Garuccio}(1995)}]{PhysRevA.52.2535}%
  \BibitemOpen
  \bibfield  {author} {\bibinfo {author} {\bibfnamefont {A.}~\bibnamefont
  {Garuccio}},\ }\href {\doibase 10.1103/PhysRevA.52.2535} {\bibfield
  {journal} {\bibinfo  {journal} {Phys. Rev. A}\ }\textbf {\bibinfo {volume}
  {52}},\ \bibinfo {pages} {2535} (\bibinfo {year} {1995})}\BibitemShut
  {NoStop}%
\bibitem [{\citenamefont {Braun}\ and\ \citenamefont
  {Choi}(2008)}]{PhysRevA.78.032114}%
  \BibitemOpen
  \bibfield  {author} {\bibinfo {author} {\bibfnamefont {D.}~\bibnamefont
  {Braun}}\ and\ \bibinfo {author} {\bibfnamefont {M.-S.}\ \bibnamefont
  {Choi}},\ }\href {\doibase 10.1103/PhysRevA.78.032114} {\bibfield  {journal}
  {\bibinfo  {journal} {Phys. Rev. A}\ }\textbf {\bibinfo {volume} {78}},\
  \bibinfo {pages} {032114} (\bibinfo {year} {2008})}\BibitemShut {NoStop}%
\bibitem [{\citenamefont {Ghirardi}\ and\ \citenamefont
  {Marinatto}(2008)}]{GHIRARDI20081982}%
  \BibitemOpen
  \bibfield  {author} {\bibinfo {author} {\bibfnamefont {G.}~\bibnamefont
  {Ghirardi}}\ and\ \bibinfo {author} {\bibfnamefont {L.}~\bibnamefont
  {Marinatto}},\ }\href {\doibase
  https://doi.org/10.1016/j.physleta.2007.11.012} {\bibfield  {journal}
  {\bibinfo  {journal} {Physics Letters A}\ }\textbf {\bibinfo {volume}
  {372}},\ \bibinfo {pages} {1982} (\bibinfo {year} {2008})}\BibitemShut
  {NoStop}%
\bibitem [{\citenamefont {van Dam}\ \emph {et~al.}(2005)\citenamefont {van
  Dam}, \citenamefont {Gill},\ and\ \citenamefont {Grunwald}}]{1468301}%
  \BibitemOpen
  \bibfield  {author} {\bibinfo {author} {\bibfnamefont {W.}~\bibnamefont {van
  Dam}}, \bibinfo {author} {\bibfnamefont {R.}~\bibnamefont {Gill}}, \ and\
  \bibinfo {author} {\bibfnamefont {P.}~\bibnamefont {Grunwald}},\ }\href
  {\doibase 10.1109/TIT.2005.851738} {\bibfield  {journal} {\bibinfo  {journal}
  {IEEE Transactions on Information Theory}\ }\textbf {\bibinfo {volume}
  {51}},\ \bibinfo {pages} {2812} (\bibinfo {year} {2005})}\BibitemShut
  {NoStop}%
\bibitem [{\citenamefont {Ac\'{\i}n}\ \emph {et~al.}(2005)\citenamefont
  {Ac\'{\i}n}, \citenamefont {Gill},\ and\ \citenamefont
  {Gisin}}]{PhysRevLett.95.210402}%
  \BibitemOpen
  \bibfield  {author} {\bibinfo {author} {\bibfnamefont {A.}~\bibnamefont
  {Ac\'{\i}n}}, \bibinfo {author} {\bibfnamefont {R.}~\bibnamefont {Gill}}, \
  and\ \bibinfo {author} {\bibfnamefont {N.}~\bibnamefont {Gisin}},\ }\href
  {\doibase 10.1103/PhysRevLett.95.210402} {\bibfield  {journal} {\bibinfo
  {journal} {Phys. Rev. Lett.}\ }\textbf {\bibinfo {volume} {95}},\ \bibinfo
  {pages} {210402} (\bibinfo {year} {2005})}\BibitemShut {NoStop}%
\bibitem [{\citenamefont {Brunner}\ \emph {et~al.}(2005)\citenamefont
  {Brunner}, \citenamefont {Gisin},\ and\ \citenamefont
  {Scarani}}]{Brunner_2005}%
  \BibitemOpen
  \bibfield  {author} {\bibinfo {author} {\bibfnamefont {N.}~\bibnamefont
  {Brunner}}, \bibinfo {author} {\bibfnamefont {N.}~\bibnamefont {Gisin}}, \
  and\ \bibinfo {author} {\bibfnamefont {V.}~\bibnamefont {Scarani}},\ }\href
  {\doibase 10.1088/1367-2630/7/1/088} {\bibfield  {journal} {\bibinfo
  {journal} {New Journal of Physics}\ }\textbf {\bibinfo {volume} {7}},\
  \bibinfo {pages} {88} (\bibinfo {year} {2005})}\BibitemShut {NoStop}%
\bibitem [{\citenamefont {Liang}\ \emph {et~al.}(2011)\citenamefont {Liang},
  \citenamefont {V\'ertesi},\ and\ \citenamefont
  {Brunner}}]{PhysRevA.83.022108}%
  \BibitemOpen
  \bibfield  {author} {\bibinfo {author} {\bibfnamefont {Y.-C.}\ \bibnamefont
  {Liang}}, \bibinfo {author} {\bibfnamefont {T.}~\bibnamefont {V\'ertesi}}, \
  and\ \bibinfo {author} {\bibfnamefont {N.}~\bibnamefont {Brunner}},\ }\href
  {\doibase 10.1103/PhysRevA.83.022108} {\bibfield  {journal} {\bibinfo
  {journal} {Phys. Rev. A}\ }\textbf {\bibinfo {volume} {83}},\ \bibinfo
  {pages} {022108} (\bibinfo {year} {2011})}\BibitemShut {NoStop}%
\bibitem [{\citenamefont {Dilley}\ and\ \citenamefont
  {Chitambar}(2018)}]{PhysRevA.97.062313}%
  \BibitemOpen
  \bibfield  {author} {\bibinfo {author} {\bibfnamefont {D.}~\bibnamefont
  {Dilley}}\ and\ \bibinfo {author} {\bibfnamefont {E.}~\bibnamefont
  {Chitambar}},\ }\href {\doibase 10.1103/PhysRevA.97.062313} {\bibfield
  {journal} {\bibinfo  {journal} {Phys. Rev. A}\ }\textbf {\bibinfo {volume}
  {97}},\ \bibinfo {pages} {062313} (\bibinfo {year} {2018})}\BibitemShut
  {NoStop}%
\bibitem [{\citenamefont {Matsukevich}\ \emph {et~al.}(2008)\citenamefont
  {Matsukevich}, \citenamefont {Maunz}, \citenamefont {Moehring}, \citenamefont
  {Olmschenk},\ and\ \citenamefont {Monroe}}]{PhysRevLett.100.150404}%
  \BibitemOpen
  \bibfield  {author} {\bibinfo {author} {\bibfnamefont {D.~N.}\ \bibnamefont
  {Matsukevich}}, \bibinfo {author} {\bibfnamefont {P.}~\bibnamefont {Maunz}},
  \bibinfo {author} {\bibfnamefont {D.~L.}\ \bibnamefont {Moehring}}, \bibinfo
  {author} {\bibfnamefont {S.}~\bibnamefont {Olmschenk}}, \ and\ \bibinfo
  {author} {\bibfnamefont {C.}~\bibnamefont {Monroe}},\ }\href {\doibase
  10.1103/PhysRevLett.100.150404} {\bibfield  {journal} {\bibinfo  {journal}
  {Phys. Rev. Lett.}\ }\textbf {\bibinfo {volume} {100}},\ \bibinfo {pages}
  {150404} (\bibinfo {year} {2008})}\BibitemShut {NoStop}%
\bibitem [{\citenamefont {Hofmann}\ \emph {et~al.}(2012)\citenamefont
  {Hofmann}, \citenamefont {Krug}, \citenamefont {Ortegel}, \citenamefont
  {Gérard}, \citenamefont {Weber}, \citenamefont {Rosenfeld},\ and\
  \citenamefont {Weinfurter}}]{doi:10.1126/science.1221856}%
  \BibitemOpen
  \bibfield  {author} {\bibinfo {author} {\bibfnamefont {J.}~\bibnamefont
  {Hofmann}}, \bibinfo {author} {\bibfnamefont {M.}~\bibnamefont {Krug}},
  \bibinfo {author} {\bibfnamefont {N.}~\bibnamefont {Ortegel}}, \bibinfo
  {author} {\bibfnamefont {L.}~\bibnamefont {Gérard}}, \bibinfo {author}
  {\bibfnamefont {M.}~\bibnamefont {Weber}}, \bibinfo {author} {\bibfnamefont
  {W.}~\bibnamefont {Rosenfeld}}, \ and\ \bibinfo {author} {\bibfnamefont
  {H.}~\bibnamefont {Weinfurter}},\ }\href {\doibase 10.1126/science.1221856}
  {\bibfield  {journal} {\bibinfo  {journal} {Science}\ }\textbf {\bibinfo
  {volume} {337}},\ \bibinfo {pages} {72} (\bibinfo {year} {2012})},\ \Eprint
  {http://arxiv.org/abs/https://www.science.org/doi/pdf/10.1126/science.1221856}
  {https://www.science.org/doi/pdf/10.1126/science.1221856} \BibitemShut
  {NoStop}%
\bibitem [{\citenamefont {Yokota}\ \emph {et~al.}(2009)\citenamefont {Yokota},
  \citenamefont {Yamamoto}, \citenamefont {Koashi},\ and\ \citenamefont
  {Imoto}}]{yokota2009direct}%
  \BibitemOpen
  \bibfield  {author} {\bibinfo {author} {\bibfnamefont {K.}~\bibnamefont
  {Yokota}}, \bibinfo {author} {\bibfnamefont {T.}~\bibnamefont {Yamamoto}},
  \bibinfo {author} {\bibfnamefont {M.}~\bibnamefont {Koashi}}, \ and\ \bibinfo
  {author} {\bibfnamefont {N.}~\bibnamefont {Imoto}},\ }\href@noop {}
  {\bibfield  {journal} {\bibinfo  {journal} {New Journal of Physics}\ }\textbf
  {\bibinfo {volume} {11}},\ \bibinfo {pages} {033011} (\bibinfo {year}
  {2009})}\BibitemShut {NoStop}%
\bibitem [{\citenamefont {Lundeen}\ and\ \citenamefont
  {Steinberg}(2009)}]{PhysRevLett.102.020404}%
  \BibitemOpen
  \bibfield  {author} {\bibinfo {author} {\bibfnamefont {J.~S.}\ \bibnamefont
  {Lundeen}}\ and\ \bibinfo {author} {\bibfnamefont {A.~M.}\ \bibnamefont
  {Steinberg}},\ }\href {\doibase 10.1103/PhysRevLett.102.020404} {\bibfield
  {journal} {\bibinfo  {journal} {Phys. Rev. Lett.}\ }\textbf {\bibinfo
  {volume} {102}},\ \bibinfo {pages} {020404} (\bibinfo {year}
  {2009})}\BibitemShut {NoStop}%
\bibitem [{\citenamefont {Luo}\ \emph {et~al.}(2018)\citenamefont {Luo},
  \citenamefont {Su}, \citenamefont {Huang}, \citenamefont {Wang},
  \citenamefont {Yang}, \citenamefont {Li}, \citenamefont {Liu}, \citenamefont
  {Chen}, \citenamefont {Lu},\ and\ \citenamefont {Pan}}]{luo2018experimental}%
  \BibitemOpen
  \bibfield  {author} {\bibinfo {author} {\bibfnamefont {Y.-H.}\ \bibnamefont
  {Luo}}, \bibinfo {author} {\bibfnamefont {H.-Y.}\ \bibnamefont {Su}},
  \bibinfo {author} {\bibfnamefont {H.-L.}\ \bibnamefont {Huang}}, \bibinfo
  {author} {\bibfnamefont {X.-L.}\ \bibnamefont {Wang}}, \bibinfo {author}
  {\bibfnamefont {T.}~\bibnamefont {Yang}}, \bibinfo {author} {\bibfnamefont
  {L.}~\bibnamefont {Li}}, \bibinfo {author} {\bibfnamefont {N.-L.}\
  \bibnamefont {Liu}}, \bibinfo {author} {\bibfnamefont {J.-L.}\ \bibnamefont
  {Chen}}, \bibinfo {author} {\bibfnamefont {C.-Y.}\ \bibnamefont {Lu}}, \ and\
  \bibinfo {author} {\bibfnamefont {J.-W.}\ \bibnamefont {Pan}},\ }\href@noop
  {} {\bibfield  {journal} {\bibinfo  {journal} {Science bulletin}\ }\textbf
  {\bibinfo {volume} {63}},\ \bibinfo {pages} {1611} (\bibinfo {year}
  {2018})}\BibitemShut {NoStop}%
\bibitem [{\citenamefont {Nguyen}\ \emph {et~al.}(2023)\citenamefont {Nguyen},
  \citenamefont {Bach}, \citenamefont {Nguyen}, \citenamefont {Tran},
  \citenamefont {Nguyen},\ and\ \citenamefont {Nguyen}}]{PhysRevD.108.023013}%
  \BibitemOpen
  \bibfield  {author} {\bibinfo {author} {\bibfnamefont {H.~C.}\ \bibnamefont
  {Nguyen}}, \bibinfo {author} {\bibfnamefont {B.~G.}\ \bibnamefont {Bach}},
  \bibinfo {author} {\bibfnamefont {T.~D.}\ \bibnamefont {Nguyen}}, \bibinfo
  {author} {\bibfnamefont {D.~M.}\ \bibnamefont {Tran}}, \bibinfo {author}
  {\bibfnamefont {D.~V.}\ \bibnamefont {Nguyen}}, \ and\ \bibinfo {author}
  {\bibfnamefont {H.~Q.}\ \bibnamefont {Nguyen}},\ }\href {\doibase
  10.1103/PhysRevD.108.023013} {\bibfield  {journal} {\bibinfo  {journal}
  {Phys. Rev. D}\ }\textbf {\bibinfo {volume} {108}},\ \bibinfo {pages}
  {023013} (\bibinfo {year} {2023})}\BibitemShut {NoStop}%
\bibitem [{\citenamefont {O'Malley}\ \emph {et~al.}(2016)\citenamefont
  {O'Malley}, \citenamefont {Babbush}, \citenamefont {Kivlichan}, \citenamefont
  {Romero}, \citenamefont {McClean}, \citenamefont {Barends}, \citenamefont
  {Kelly}, \citenamefont {Roushan}, \citenamefont {Tranter}, \citenamefont
  {Ding}, \citenamefont {Campbell}, \citenamefont {Chen}, \citenamefont {Chen},
  \citenamefont {Chiaro}, \citenamefont {Dunsworth}, \citenamefont {Fowler},
  \citenamefont {Jeffrey}, \citenamefont {Lucero}, \citenamefont {Megrant},
  \citenamefont {Mutus}, \citenamefont {Neeley}, \citenamefont {Neill},
  \citenamefont {Quintana}, \citenamefont {Sank}, \citenamefont {Vainsencher},
  \citenamefont {Wenner}, \citenamefont {White}, \citenamefont {Coveney},
  \citenamefont {Love}, \citenamefont {Neven}, \citenamefont {Aspuru-Guzik},\
  and\ \citenamefont {Martinis}}]{PhysRevX.6.031007}%
  \BibitemOpen
  \bibfield  {author} {\bibinfo {author} {\bibfnamefont {P.~J.~J.}\
  \bibnamefont {O'Malley}}, \bibinfo {author} {\bibfnamefont {R.}~\bibnamefont
  {Babbush}}, \bibinfo {author} {\bibfnamefont {I.~D.}\ \bibnamefont
  {Kivlichan}}, \bibinfo {author} {\bibfnamefont {J.}~\bibnamefont {Romero}},
  \bibinfo {author} {\bibfnamefont {J.~R.}\ \bibnamefont {McClean}}, \bibinfo
  {author} {\bibfnamefont {R.}~\bibnamefont {Barends}}, \bibinfo {author}
  {\bibfnamefont {J.}~\bibnamefont {Kelly}}, \bibinfo {author} {\bibfnamefont
  {P.}~\bibnamefont {Roushan}}, \bibinfo {author} {\bibfnamefont
  {A.}~\bibnamefont {Tranter}}, \bibinfo {author} {\bibfnamefont
  {N.}~\bibnamefont {Ding}}, \bibinfo {author} {\bibfnamefont {B.}~\bibnamefont
  {Campbell}}, \bibinfo {author} {\bibfnamefont {Y.}~\bibnamefont {Chen}},
  \bibinfo {author} {\bibfnamefont {Z.}~\bibnamefont {Chen}}, \bibinfo {author}
  {\bibfnamefont {B.}~\bibnamefont {Chiaro}}, \bibinfo {author} {\bibfnamefont
  {A.}~\bibnamefont {Dunsworth}}, \bibinfo {author} {\bibfnamefont {A.~G.}\
  \bibnamefont {Fowler}}, \bibinfo {author} {\bibfnamefont {E.}~\bibnamefont
  {Jeffrey}}, \bibinfo {author} {\bibfnamefont {E.}~\bibnamefont {Lucero}},
  \bibinfo {author} {\bibfnamefont {A.}~\bibnamefont {Megrant}}, \bibinfo
  {author} {\bibfnamefont {J.~Y.}\ \bibnamefont {Mutus}}, \bibinfo {author}
  {\bibfnamefont {M.}~\bibnamefont {Neeley}}, \bibinfo {author} {\bibfnamefont
  {C.}~\bibnamefont {Neill}}, \bibinfo {author} {\bibfnamefont
  {C.}~\bibnamefont {Quintana}}, \bibinfo {author} {\bibfnamefont
  {D.}~\bibnamefont {Sank}}, \bibinfo {author} {\bibfnamefont {A.}~\bibnamefont
  {Vainsencher}}, \bibinfo {author} {\bibfnamefont {J.}~\bibnamefont {Wenner}},
  \bibinfo {author} {\bibfnamefont {T.~C.}\ \bibnamefont {White}}, \bibinfo
  {author} {\bibfnamefont {P.~V.}\ \bibnamefont {Coveney}}, \bibinfo {author}
  {\bibfnamefont {P.~J.}\ \bibnamefont {Love}}, \bibinfo {author}
  {\bibfnamefont {H.}~\bibnamefont {Neven}}, \bibinfo {author} {\bibfnamefont
  {A.}~\bibnamefont {Aspuru-Guzik}}, \ and\ \bibinfo {author} {\bibfnamefont
  {J.~M.}\ \bibnamefont {Martinis}},\ }\href {\doibase
  10.1103/PhysRevX.6.031007} {\bibfield  {journal} {\bibinfo  {journal} {Phys.
  Rev. X}\ }\textbf {\bibinfo {volume} {6}},\ \bibinfo {pages} {031007}
  (\bibinfo {year} {2016})}\BibitemShut {NoStop}%
\bibitem [{\citenamefont {Haah}\ \emph {et~al.}(2023)\citenamefont {Haah},
  \citenamefont {Hastings}, \citenamefont {Kothari},\ and\ \citenamefont
  {Low}}]{doi:10.1137/18M1231511}%
  \BibitemOpen
  \bibfield  {author} {\bibinfo {author} {\bibfnamefont {J.}~\bibnamefont
  {Haah}}, \bibinfo {author} {\bibfnamefont {M.~B.}\ \bibnamefont {Hastings}},
  \bibinfo {author} {\bibfnamefont {R.}~\bibnamefont {Kothari}}, \ and\
  \bibinfo {author} {\bibfnamefont {G.~H.}\ \bibnamefont {Low}},\ }\href
  {\doibase 10.1137/18M1231511} {\bibfield  {journal} {\bibinfo  {journal}
  {SIAM Journal on Computing}\ }\textbf {\bibinfo {volume} {52}},\ \bibinfo
  {pages} {FOCS18} (\bibinfo {year} {2023})},\ \Eprint
  {http://arxiv.org/abs/https://doi.org/10.1137/18M1231511}
  {https://doi.org/10.1137/18M1231511} \BibitemShut {NoStop}%
\bibitem [{\citenamefont {Hempel}\ \emph {et~al.}(2018)\citenamefont {Hempel},
  \citenamefont {Maier}, \citenamefont {Romero}, \citenamefont {McClean},
  \citenamefont {Monz}, \citenamefont {Shen}, \citenamefont {Jurcevic},
  \citenamefont {Lanyon}, \citenamefont {Love}, \citenamefont {Babbush},
  \citenamefont {Aspuru-Guzik}, \citenamefont {Blatt},\ and\ \citenamefont
  {Roos}}]{PhysRevX.8.031022}%
  \BibitemOpen
  \bibfield  {author} {\bibinfo {author} {\bibfnamefont {C.}~\bibnamefont
  {Hempel}}, \bibinfo {author} {\bibfnamefont {C.}~\bibnamefont {Maier}},
  \bibinfo {author} {\bibfnamefont {J.}~\bibnamefont {Romero}}, \bibinfo
  {author} {\bibfnamefont {J.}~\bibnamefont {McClean}}, \bibinfo {author}
  {\bibfnamefont {T.}~\bibnamefont {Monz}}, \bibinfo {author} {\bibfnamefont
  {H.}~\bibnamefont {Shen}}, \bibinfo {author} {\bibfnamefont {P.}~\bibnamefont
  {Jurcevic}}, \bibinfo {author} {\bibfnamefont {B.~P.}\ \bibnamefont
  {Lanyon}}, \bibinfo {author} {\bibfnamefont {P.}~\bibnamefont {Love}},
  \bibinfo {author} {\bibfnamefont {R.}~\bibnamefont {Babbush}}, \bibinfo
  {author} {\bibfnamefont {A.}~\bibnamefont {Aspuru-Guzik}}, \bibinfo {author}
  {\bibfnamefont {R.}~\bibnamefont {Blatt}}, \ and\ \bibinfo {author}
  {\bibfnamefont {C.~F.}\ \bibnamefont {Roos}},\ }\href {\doibase
  10.1103/PhysRevX.8.031022} {\bibfield  {journal} {\bibinfo  {journal} {Phys.
  Rev. X}\ }\textbf {\bibinfo {volume} {8}},\ \bibinfo {pages} {031022}
  (\bibinfo {year} {2018})}\BibitemShut {NoStop}%
\bibitem [{\citenamefont {Kokail}\ \emph {et~al.}(2019)\citenamefont {Kokail},
  \citenamefont {Maier}, \citenamefont {van Bijnen}, \citenamefont {Brydges},
  \citenamefont {Joshi}, \citenamefont {Jurcevic}, \citenamefont {Muschik},
  \citenamefont {Silvi}, \citenamefont {Blatt}, \citenamefont {Roos} \emph
  {et~al.}}]{kokail2019self}%
  \BibitemOpen
  \bibfield  {author} {\bibinfo {author} {\bibfnamefont {C.}~\bibnamefont
  {Kokail}}, \bibinfo {author} {\bibfnamefont {C.}~\bibnamefont {Maier}},
  \bibinfo {author} {\bibfnamefont {R.}~\bibnamefont {van Bijnen}}, \bibinfo
  {author} {\bibfnamefont {T.}~\bibnamefont {Brydges}}, \bibinfo {author}
  {\bibfnamefont {M.~K.}\ \bibnamefont {Joshi}}, \bibinfo {author}
  {\bibfnamefont {P.}~\bibnamefont {Jurcevic}}, \bibinfo {author}
  {\bibfnamefont {C.~A.}\ \bibnamefont {Muschik}}, \bibinfo {author}
  {\bibfnamefont {P.}~\bibnamefont {Silvi}}, \bibinfo {author} {\bibfnamefont
  {R.}~\bibnamefont {Blatt}}, \bibinfo {author} {\bibfnamefont {C.~F.}\
  \bibnamefont {Roos}},  \emph {et~al.},\ }\href@noop {} {\bibfield  {journal}
  {\bibinfo  {journal} {Nature}\ }\textbf {\bibinfo {volume} {569}},\ \bibinfo
  {pages} {355} (\bibinfo {year} {2019})}\BibitemShut {NoStop}%
\bibitem [{\citenamefont {Martinez}\ \emph {et~al.}(2016)\citenamefont
  {Martinez}, \citenamefont {Muschik}, \citenamefont {Schindler}, \citenamefont
  {Nigg}, \citenamefont {Erhard}, \citenamefont {Heyl}, \citenamefont {Hauke},
  \citenamefont {Dalmonte}, \citenamefont {Monz}, \citenamefont {Zoller} \emph
  {et~al.}}]{martinez2016real}%
  \BibitemOpen
  \bibfield  {author} {\bibinfo {author} {\bibfnamefont {E.~A.}\ \bibnamefont
  {Martinez}}, \bibinfo {author} {\bibfnamefont {C.~A.}\ \bibnamefont
  {Muschik}}, \bibinfo {author} {\bibfnamefont {P.}~\bibnamefont {Schindler}},
  \bibinfo {author} {\bibfnamefont {D.}~\bibnamefont {Nigg}}, \bibinfo {author}
  {\bibfnamefont {A.}~\bibnamefont {Erhard}}, \bibinfo {author} {\bibfnamefont
  {M.}~\bibnamefont {Heyl}}, \bibinfo {author} {\bibfnamefont {P.}~\bibnamefont
  {Hauke}}, \bibinfo {author} {\bibfnamefont {M.}~\bibnamefont {Dalmonte}},
  \bibinfo {author} {\bibfnamefont {T.}~\bibnamefont {Monz}}, \bibinfo {author}
  {\bibfnamefont {P.}~\bibnamefont {Zoller}},  \emph {et~al.},\ }\href@noop {}
  {\bibfield  {journal} {\bibinfo  {journal} {Nature}\ }\textbf {\bibinfo
  {volume} {534}},\ \bibinfo {pages} {516} (\bibinfo {year}
  {2016})}\BibitemShut {NoStop}%
\bibitem [{\citenamefont {Havl{\'\i}{\v{c}}ek}\ \emph
  {et~al.}(2019)\citenamefont {Havl{\'\i}{\v{c}}ek}, \citenamefont
  {C{\'o}rcoles}, \citenamefont {Temme}, \citenamefont {Harrow}, \citenamefont
  {Kandala}, \citenamefont {Chow},\ and\ \citenamefont
  {Gambetta}}]{havlivcek2019supervised}%
  \BibitemOpen
  \bibfield  {author} {\bibinfo {author} {\bibfnamefont {V.}~\bibnamefont
  {Havl{\'\i}{\v{c}}ek}}, \bibinfo {author} {\bibfnamefont {A.~D.}\
  \bibnamefont {C{\'o}rcoles}}, \bibinfo {author} {\bibfnamefont
  {K.}~\bibnamefont {Temme}}, \bibinfo {author} {\bibfnamefont {A.~W.}\
  \bibnamefont {Harrow}}, \bibinfo {author} {\bibfnamefont {A.}~\bibnamefont
  {Kandala}}, \bibinfo {author} {\bibfnamefont {J.~M.}\ \bibnamefont {Chow}}, \
  and\ \bibinfo {author} {\bibfnamefont {J.~M.}\ \bibnamefont {Gambetta}},\
  }\href@noop {} {\bibfield  {journal} {\bibinfo  {journal} {Nature}\ }\textbf
  {\bibinfo {volume} {567}},\ \bibinfo {pages} {209} (\bibinfo {year}
  {2019})}\BibitemShut {NoStop}%
\bibitem [{\citenamefont {Cong}\ \emph {et~al.}(2019)\citenamefont {Cong},
  \citenamefont {Choi},\ and\ \citenamefont {Lukin}}]{cong2019quantum}%
  \BibitemOpen
  \bibfield  {author} {\bibinfo {author} {\bibfnamefont {I.}~\bibnamefont
  {Cong}}, \bibinfo {author} {\bibfnamefont {S.}~\bibnamefont {Choi}}, \ and\
  \bibinfo {author} {\bibfnamefont {M.~D.}\ \bibnamefont {Lukin}},\ }\href@noop
  {} {\bibfield  {journal} {\bibinfo  {journal} {Nature Physics}\ }\textbf
  {\bibinfo {volume} {15}},\ \bibinfo {pages} {1273} (\bibinfo {year}
  {2019})}\BibitemShut {NoStop}%
\bibitem [{\citenamefont {Farhi}\ and\ \citenamefont
  {Neven}(2018)}]{farhi2018classification}%
  \BibitemOpen
  \bibfield  {author} {\bibinfo {author} {\bibfnamefont {E.}~\bibnamefont
  {Farhi}}\ and\ \bibinfo {author} {\bibfnamefont {H.}~\bibnamefont {Neven}},\
  }\href@noop {} {\bibfield  {journal} {\bibinfo  {journal} {arXiv preprint
  arXiv:1802.06002}\ } (\bibinfo {year} {2018})}\BibitemShut {NoStop}%
\end{thebibliography}%

\newpage
\appendix

\section{Identifying the cause of discrepancies between Brisbane measurement and simulation}

In this section, we'll discuss about experiments designed to identify the causes of deviations in Brisbane measurement results. Each experiment will have two data sets: one from IBM Brisbane and one from the simulation. The first two error sources that we want to examine are the number of qubits and the number of Toffoli gates. The factors may contribute to deviations in results, but we want to determine whether their influence was considerable.

\begin{figure}[ht]
    \centering
    \includegraphics[width=1\linewidth]{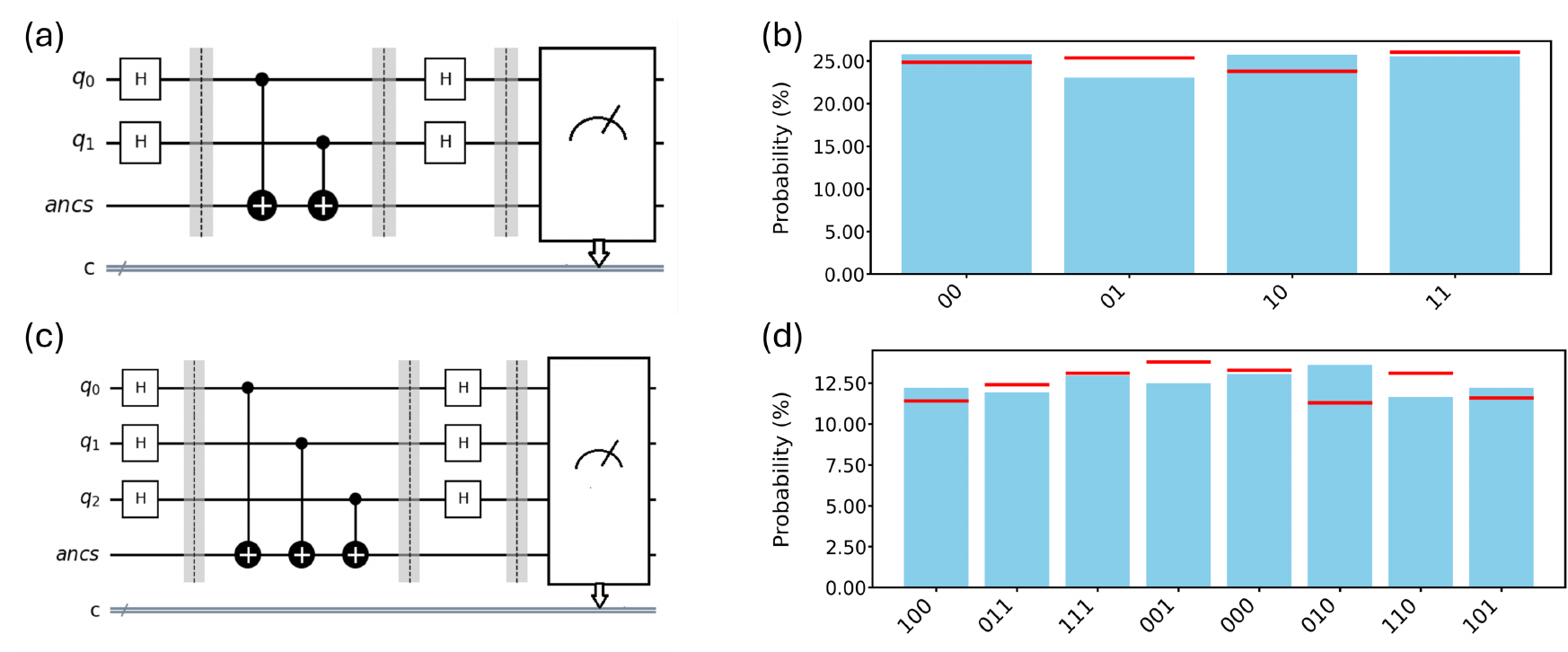}
    \caption{Experiments used to observe the impact of the number of qubits on measurement results. To avoid the effect of $\theta$, the quantum gates used in these experiments are only Hadamard gates. Two Hadamard gates will be applied to each qubit. Additionally, each qubit controls an individual ancilla qubit. The blue columns are results obtained from IBM Brisbane, meanwhile, the red horizontal lines are simulation results. (a, b) Quantum circuit used to examine two qubits and the corresponding measurement results. (c, d) Quantum circuit used to examine three qubits and the corresponding measurement results.}
    \label{6}
\end{figure}

Fig. \ref{6} shows two experiments that are used to answer this question. The main target of these two experiments is to verify the impact of the number of qubits and Toffoli gates on measurement results, so we decided to use the Hadamard gates to avoid the impact of $\theta$ on our observation. Furthermore, we configured the Toffoli gates with only one control, one target qubit, and no overlapping qubits among the gates. This simplified configuration allows us to focus primarily on the number of gates. With this setup, the results of the two circuits (6a) and (6c) mostly depend on the number of qubits and the number of Toffoli gates. Fig. \ref{6}(b) shows the measurement results of Fig. \ref{6}(a) for the case of two qubits. The deviations between Brisbane results and their simulation calculations are less than 3\%, which is acceptable. Moving to the three-qubit circuit Fig. \ref{6}(c), the corresponding results depicted in Fig. \ref{6}(d) also show deviations of Bribsane and the simulation below 4\%. These deviations are not only acceptable but also nearly equal to the deviation of the two-qubit case. These results suggest that while the errors may still occur, the deviations are acceptable, and the number of qubits and Toffoli gates may not introduce significant discrepancies. The deviations remain consistent and acceptable even as the number of these elements increases. Therefore, from these results, we assume that the number of qubits and Toffoli gates may not be the major reason for making considerable deviations.

\begin{figure}[H]
    \centering
    \includegraphics[width=1\linewidth]{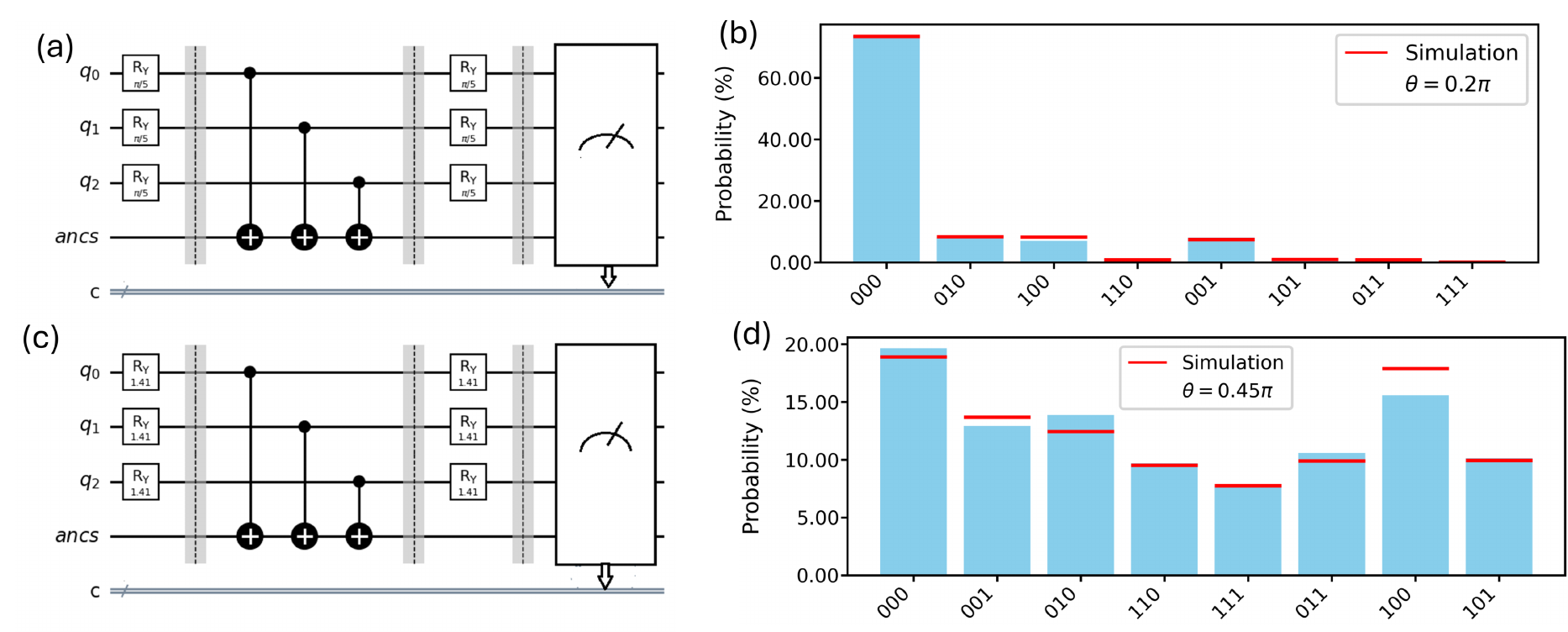}
    \caption{Experiments used to observe the impact of $\theta$ value on measurement results. The circuits' structure are the same as Fig. 6c, only replacing the Hadamard gates with $R_y(\theta)$ gates. Each qubit still controls an individual ancilla qubit, and no overlapping control qubits among Toffoli gates. The blue columns are results obtained from IBM Brisbane; meanwhile, the red horizontal lines are simulation results. (a, b) Quantum circuit used to examine $\theta=0.2\pi(rad)$ and the corresponding measurement results. (c, d) Quantum circuit used to examine $\theta=0.45\pi(rad)$ and the corresponding measurement results.}
    \label{7}
\end{figure}

The next factor to examine is the value of $\theta$ used in $R_y$ gates. Two circuits are used in this examination. Each contains three qubits and shares the same structure as Fig. \ref{6}c, except the Hadamard gates are now replaced by $R_y(\theta)$ gates. We considered two values of $\theta$, which are $0.2\pi(rad)$ and $0.45\pi(rad)$. Fig. \ref{7} illustrates circuits and their corresponding results for this $\theta$ examination. The first value of $\theta$ under examination, $0.2\pi(rad)$, was chosen because it is sufficiently small to observe the difference in results with larger values. Fig. \ref{7}(a) is the circuit using $R_y(\theta=0.2\pi)$, and its corresponding results are depicted in Fig. \ref{7}(b). The outcome for $\theta=0.2\pi(rad)$ is encouraging as the deviations between Brisbane and the simulation results are minimal. However, when $\theta$ is increased to $0.45\pi(rad)$, the discrepancies start to emerge. Fig. \ref{7}(c) is the circuit using $R_y(\theta=0.45\pi)$ gates, and Fig. \ref{7}(d) is its corresponding results. As shown in Fig. \ref{7}(d), the deviations start to increase, rising from approximately below 1\% for $\theta=0.2\pi(rad)$ to approximately 4\% for $\theta=0.2\pi(rad)$. While these deviations remain within acceptable limits, this trend suggests that increasing $\theta$ values may correlate with increasing discrepancies. Therefore, we assume that the $\theta$ value may not be the main reason for making significant discrepancies, but it could exacerbate deviations.

\begin{figure}[H]
    \centering
    \includegraphics[width=1\linewidth]{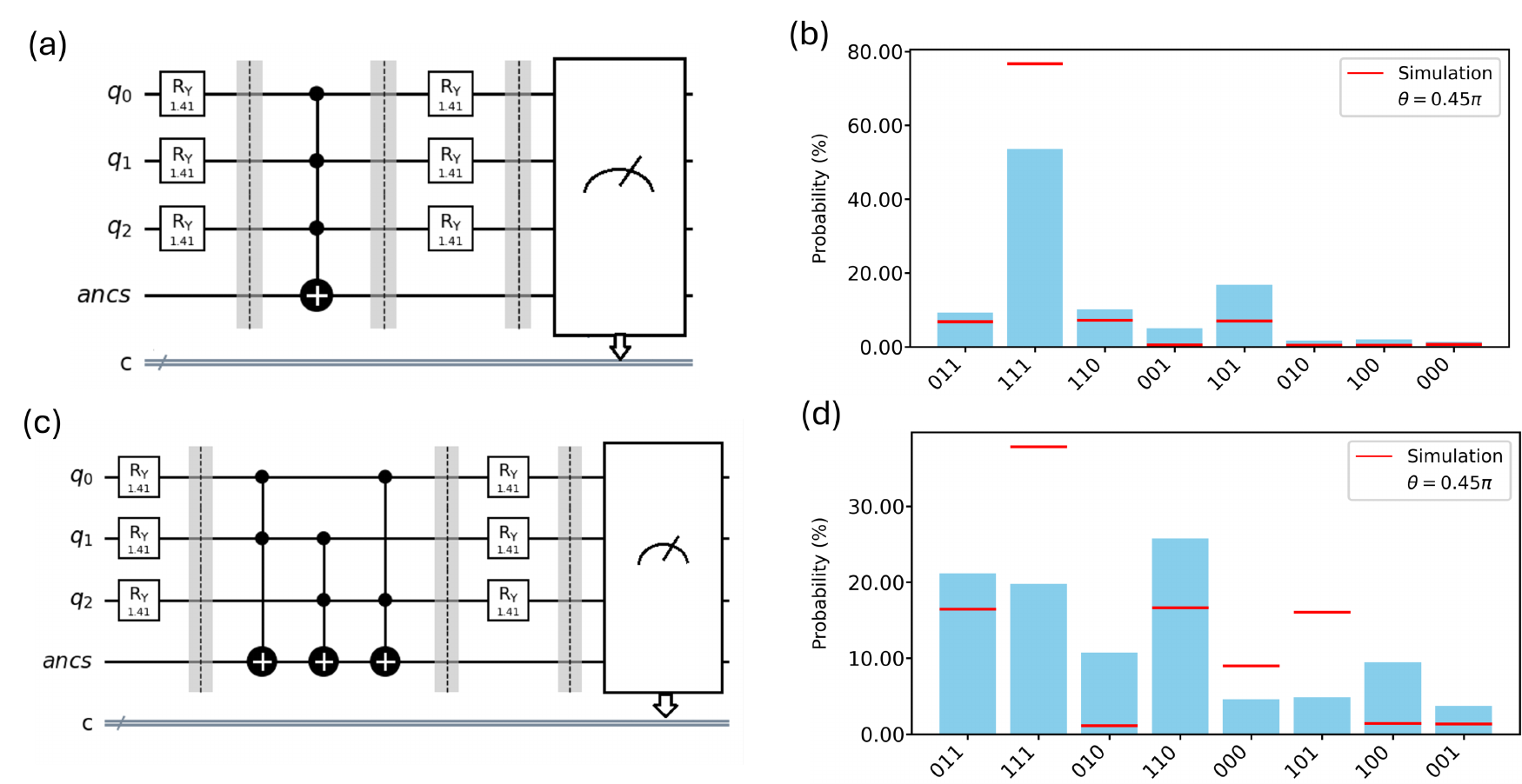}
    \caption{Experiments used to observe the impact of the number of control qubits on measurement results. Each circuit has three qubits, and both of them use $R_y(\theta=0.45\pi)$. The number of control and target qubits in each circuit is now different. The blue columns are results obtained from IBM Brisbane; meanwhile, the red horizontal lines are simulation results. (a, b) Quantum circuit used to examine and the corresponding measurement results. Only one Toffoli gate is used in this circuit. The number of control qubits is three, and only one target qubit, which is the ancilla. (c, d) Quantum circuit used to examine and the corresponding measurement results. This circuit has three Toffoli gates, each of which has two control qubits. Each pair controls an individual ancilla qubit. Additionally, there are overlapping control qubits in this circuit.}
    \label{8}
\end{figure}

The last two aspects that we investigated are the number of control qubits in a Toffoli gate and the existence of overlapping control qubits among these gates. These two are characteristic of circuits used in the three sets of conditions. Two circuits were designed for this examination, each has three qubits, and the $R_y$ gates use $\theta=0.45\pi(rad)$, but differ in the number of control and target qubits. Specifically, the circuit shown in Fig. \ref{8}(a) only has one Toffoli gate, where all three qubits control a single ancilla. Whereas the circuit shown in Fig. \ref{8}(c) has three Toffoli gates and three distinct ancillas. Each gate uses two qubits to control an individual ancilla, and overlapping control qubits are present among these gates. For instance, in Fig. \ref{8}(c), qubit $q_1$ participates as a control qubit in two different Toffoli gates, once in a pair with $q_0$ and again in a pair with $q_2$. Fig. \ref{8}(b) shows the corresponding results for the circuit (8a), and reveals significant deviations. The largest deviation belongs to $\ket{111}$ reaching up to more than 20\%. Similarly, Fig. \ref{8}(d) shows the results for the circuit (8c), and the difference between the Brisbane results and their simulation results remains substantial. The largest deviation is also up to 20\%. From the results of two earlier experiments shown in Fig. \ref{6} - \ref{7}, where the number of Toffoli gates is also three, the deviations were much smaller, which are less than 4\%. The difference is that in the earlier experiments, each Toffoli gate had only one control qubit, and there were no overlapping qubits among them. Based on these findings, we assume that if the number of control qubits is more than one in a Toffoli gate, and there are overlapping qubits among these Toffoli gates may cause considerable deviations.

In conclusion, we assume that the number of qubits and Toffoli gates may not be the main reasons for causing considerable deviation between Brisbane results and the simulation calculations. The major reason contributing to these discrepancies might be the number of control qubits used in a Toffoli gate, along with the presence of overlapping control qubits among Toffoli gates. Additionally, the value of $\theta$ used in $R_y$ gates also contributes to the deviations, specifically, it may exacerbate the deviations.

\section{Normalization constant}

Using the state equation Eq.\eqref{e30}, the normalization constant N is calculated as follows
\begin{equation}
    \begin{split}
        \langle \psi | \psi \rangle&=N^2[B_1^*A_2^*B_3^*A_4^*\bra{v_1u_2v_3u_4}+B_1^*B_2^*B_3^*A_4^*\bra{v_1v_2v_3u_4}+A_1^*B_2^*A_3^*B_4^*\bra{u_1v_2u_3v_4}\\&\quad+B_1^*B_2^*A_3^*B_4^*\bra{v_1v_2u_3v_4}+A_1^*B_2^*B_3^*B_4^*\bra{u_1v_2v_3v_4}+B_1^*A_2^*B_3^*B_4^*\bra{v_1u_2v_3v_4}\\&\quad+B_1^*B_2^*B_3^*B_4^*\bra{v_1v_2v_3v_4}]\times[B_1A_2B_3A_4\ket{v_1u_2v_3u_4}+B_1B_2B_3A_4\ket{v_1v_2v_3u_4}\\&\quad+A_1B_2A_3B_4\ket{u_1v_2u_3v_4}+B_1B_2A_3B_4\ket{v_1v_2u_3v_4}+A_1B_2B_3B_4\ket{u_1v_2v_3v_4}\\&\quad+B_1A_2B_3B_4\ket{v_1u_2v_3v_4}+B_1B_2B_3B_4\ket{v_1v_2v_3v_4}],
        \label{app1}
    \end{split}
\end{equation}
\begin{equation}
    \begin{split}
        \Rightarrow\langle \psi | \psi \rangle&=N^2[|B_1|^2|A_2|^2|B_3|^2|A_4|^2+|B_1|^2|B_2|^2|B_3|^2|A_4|^2+|A_1|^2|B_2|^2|A_3|^2|B_4|^2\\&\quad+|B_1|^2|B_2|^2|A_3|^2|B_4|^2+|A_1|^2|B_2|^2|B_3|^2|B_4|^2+|B_1|^2|A_2|^2|B_3|^2|B_4|^2\\&\quad+|B_1|^2|B_2|^2|B_3|^2|B_4|^2]\\
        &=N^2[1-|A_1|^2|A_2|^2|A_3|^2|A_4|^2-|A_1|^2|A_2|^2|A_3|^2|B_4|^2-|A_1|^2|A_2|^2|B_3|^2|A_4|^2\\&\quad-|A_1|^2|A_2|^2|B_3|^2|B_4|^2-|B_1|^2|A_2|^2|A_3|^2|B_4|^2-|B_1|^2|A_2|^2|A_3|^2|A_4|^2\\&\quad-|B_1|^2|B_2|^2|A_3|^2|A_4|^2-|A_1|^2|B_2|^2|A_3|^2|A_4|^2-|A_1|^2|B_2|^2|B_3|^2|A_4|^2],
        \label{app2}
    \end{split}
\end{equation}
where the part
\begin{equation}
    \begin{split}
        &-|A_1|^2|A_2|^2|A_3|^2|A_4|^2-|A_1|^2|A_2|^2|A_3|^2|B_4|^2-|A_1|^2|A_2|^2|B_3|^2|A_4|^2-|A_1|^2|A_2|^2|B_3|^2|B_4|^2\\&-|B_1|^2|A_2|^2|A_3|^2|B_4|^2-|B_1|^2|A_2|^2|A_3|^2|A_4|^2-|B_1|^2|B_2|^2|A_3|^2|A_4|^2-|A_1|^2|B_2|^2|A_3|^2|A_4|^2\\&-|A_1|^2|B_2|^2|B_3|^2|A_4|^2,
    \end{split}
\end{equation}
can be rewritten as C in Eq.\eqref{e31}. Therefore Eq. \eqref{app2} can be rewritten as
\begin{equation}
    \langle \psi | \psi \rangle=N^2[1+C].
\end{equation}
From the normalization condition $\langle \psi | \psi \rangle=1$, we obtain $$N=\frac{1}{\sqrt{1+C}}$$.

\end{document}